\PassOptionsToPackage{hyphens}{url}
\documentclass[pdflatex,sn-basic]{sn-jnl}

\usepackage{kotex}
\usepackage{graphicx}
\usepackage{multirow}
\usepackage{amsmath,amssymb,amsfonts}
\usepackage{booktabs}
\usepackage{xcolor}
\usepackage{textcomp}
\usepackage{manyfoot}
\usepackage{placeins}
\usepackage{algorithm}
\usepackage{algorithmicx}
\usepackage{algpseudocode}
\usepackage{listings}

  \tolerance=2000 \hyphenpenalty=7000 \exhyphenpenalty=7000 \hbadness=10000 \vbadness=10000 \hfuzz=2pt \vfuzz=2pt \hypersetup{hypertexnames=false}

\begin{document}

\title[NPK Audit against Official References]{Marginal Alignment Does Not Guarantee Joint-Distribution Fidelity: An Official-Reference Audit of Nemotron-Personas-Korea with Cross-Locale Replication}

\author*[1]{\fnm{Joonhyung} \sur{Bae}}\email{jh.bae@kaist.ac.kr}
\affil[1]{\orgname{Korea Advanced Institute of Science and Technology (KAIST)}, \city{Daejeon}, \country{South Korea}; ORCID: 0000-0001-5933-4302}

\abstract{Synthetic persona datasets cite alignment with official demographics as a basis for trust, yet downstream users consume them as joint structures across age, sex, region, occupation, education, name, and institutional status. Marginal alignment does not imply that these joints are preserved. We propose the Independence-Assumption Footprint (IAF), an audit primitive that operates on the attribute combinations a dataset card itself documents as treated independently. For each such combination, IAF compares the synthetic joint against an external official or institutional reference, using direct joint tables where available and rule-implied checks otherwise. Applied to NVIDIA Nemotron-Personas-Korea (one million Korean synthetic personas), IAF finds that NPK aligns with KOSIS marginals while three joints fail. The major-by-occupation distribution against the KEIS graduate universe carries a large conditional mismatch. The age profile of military service is institutionally inconsistent. Female representation in male-dominated occupations is substantially over-flattened toward parity, with the strict screening verdict mapping-dependent and age-robust under direct standardisation. A transferability demonstration across six further NPK locales finds locale-dependent rather than universal diagnostics, with reference-taxonomy cardinality confounding cross-locale flag counts. For synthetic personas used as silicon samples, marginal claims must therefore be paired with disclosure-anchored joint audits before reuse. The released audit artefacts (reference manifests, occupational crosswalks, derived metrics, reproducibility scripts) instantiate this protocol on the NPK family and are released for retargeting at other synthetic persona resources.}


\maketitle

\section{Introduction}
\label{sec:intro}

Synthetic persona datasets are proliferating as silicon-sample substitutes for survey respondents \citep{argyle2023silicon}, as probes of opinion distributions \citep{santurkar2023whoseopinions}, and as fixtures in LLM evaluation \citep{consistencyai2025}. These resources typically gain credibility through claims of alignment with national demographics, but single-attribute marginal matching does not imply that the joint structure inside personas is preserved. Downstream users consume personas in which age, sex, region, occupation, education, name, and institutional status appear together, so the evaluation question for synthetic personas as a language resource is not whether the marginals look Korean but whether the verifiable joint structure of Korean society is preserved.

Korean is a particularly well-instrumented locale for this audit question. The Korean Statistical Information Service (KOSIS) publishes occupation-by-sex contingency tables at major-class resolution from the Economically Active Population Survey, the Korea Employment Information Service (KEIS) releases the Graduates Occupational Mobility Survey (GOMS) microdata as a graduate-cohort major-by-occupation reference, the Supreme Court of Korea publishes birth-year name-cohort statistics through its e-family service, and the Military Manpower Administration releases conscription cohort counts. The Korean Standard Classification of Occupations (KSCO) ties to ISCO-08 and provides a stable target taxonomy for free-text occupation labels. Together these references make a public-only joint-structure audit feasible at a granularity rarely attainable for synthetic persona resources in other languages, while the Korean labour market's well-known occupational sex segregation and male-conscription life course supply specific institutional patterns that any faithful Korean synthetic population should preserve.

Nemotron-Personas-Korea (NPK), released by NVIDIA on 2026-04-20 under CC~BY~4.0, is a well-suited test case. NPK comprises one million records and $7$ narrative fields and cites the Korean Statistical Information Service (KOSIS), the Korean Standard Classification of Occupations (KSCO), and other Korean public sources as grounding sources \citep{nvidia2026npk}, while explicitly stating that in occupation assignment, sex, income, education, and major field operate independently and interaction effects are not modelled. NPK therefore simultaneously claims alignment with official statistics and documents generative assumptions that may weaken joint structures important to the Korean labour market and life course.

We address this tension through a third-party resource validation primitive. The \emph{Independence-Assumption Footprint} (IAF) contrasts attribute combinations the dataset card documents as treated independently against external official or institutional references (KOSIS, Supreme Court of Korea name statistics, the Korea Employment Information Service (KEIS) Graduates Occupational Mobility Survey (GOMS), and Korean military service constraints). Contrasting NPK-Korea against six externally anchored axes, we report that the demographic spine is preserved while occupation-linked joints governed by the disclosed independence assumption show major$\times$occupation fidelity loss and sex$\times$KSCO attenuation/parity deviations.

We adapt the same reference-ratio diagnostic to seven NPK locales as a transferability demonstration, finding locale-dependent rather than universal diagnostics. KoBBQ survey aggregates and LoRA/downstream benchmarks are not evidence at the same level and are reported only as cultural-layer and use-scenario boundary checks. The released audit artefacts (reference-table manifests, KSCO crosswalks, derived metric tables, LLM-assist label caches, and figure-generation scripts) instantiate the validation protocol on the NPK family and are released for retargeting at other synthetic persona resources.

The paper makes three contributions. First, it introduces IAF as a disclosure-anchored audit primitive that operationalises the marginal-vs-joint gap a synthetic persona resource may leave behind. Second, it instantiates the protocol on NPK-Korea across six externally anchored axes with two pre-specified negative controls, reports the structural findings, and releases the audit artefacts under an MIT licence for retargeting. Third, it transfers the diagnostic across six further NPK locales and one non-NPK resource (PersonaHub) as a portability demonstration, noting where the diagnostic is grain-confounded or where the target resource's card does not expose the independence axes the primitive needs.

\section{Related Work}
\label{sec:related}

\subsection{Resource documentation and synthetic-persona validity}
Schemas for resource documentation include datasheets, model/data cards \citep{gebru2021datasheets,mitchell2019modelcards,pushkarna2022datacards,hutchinson2021accountability}, and Data Statements for language resources \citep{bender2018datastatements}; \citet{belgodere2023auditing} formalise a trust trade-off across fidelity, utility, privacy, fairness, and robustness, of which we close fidelity and fairness here. Large-scale LRE validation often cannot support full blind multi-annotator labelling \citep{rigoutsterryn2020term,shardlow2022complex}; the human-label-variation perspective \citep{plank2022humanlabel} and construct-validity tradition \citep{cronbach1955construct} motivate treating KSCO review as documented adjudication of mapping assumptions rather than as a new reference annotation layer. Silicon-sample validity conditions \citep{argyle2023silicon}, opinion-distribution alignment \citep{santurkar2023whoseopinions}, statistical replication tests \citep{aher2023llmsim}, and generative-agent simulators \citep{park2023generativeagents} establish the question of whom synthetic personas represent; recent work shows that LLM synthetic respondents systematically misportray demographic groups \citep{bisbee2024synthetic,dominguezolmedo2024questioning} and that persona pipelines amplify and flatten social biases \citep{li2025promise,personalitytrap2026,kumar2024subtle,rooein2025biasedtales,araujo2025persistent,batzner2025whose,ge2024personahub}. \citet{hu2025popaligned} evaluate the global Nemotron-Personas family on World Values Survey attitude alignment, complementary to our disclosure-anchored official-reference audit at the joint-distribution level. NPK's over-feminisation of male-dominated occupations is a generative-side manifestation of the bias-amplification lineage \citep{zhao2017menalsolike,bianchi2023easilyaccessible} and is compatible with model collapse \citep{shumailov2024modelcollapse} and RLHF-induced diversity reduction \citep{kirk2024rlhf} as candidate mechanisms.

\subsection{Audit primitives, LLM judges, and Korean NLP context}
Demographic parity, disparate impact, and calibration anchor the classical fairness literature \citep{hardt2016equalopp,feldman2015disparate,chouldechova2017fairpred}; NPK's selective joint-structure attenuation falls in the representation- and aggregation-bias positions of the \citet{suresh2021framework} ML-lifecycle taxonomy. Our joint-distribution analysis uses df-aware Cramér's V \citep{bergsma2013cramer} and the decomposition perspective of \citet{cheng2023compost}, with persona-bias audit precedents from \citet{cao2026pba}. BBQ \citep{parrish2022bbq} is the parent of KoBBQ \citep{jin2023kobbq}, and GlobalOpinionQA \citep{durmus2024globalopinion} is a methodological cousin of our cross-locale replication. The Korean NLP resource family \citep{park2021klue,son2024kmmlu,son2023haerae,lee2023kosbi,jeong2022kold,jin2023kobbq,song2026kolla} sets the venue context. For KSCO crosswalk uncertainty we use the panel-of-judges design of \citet{verga2024juries,zheng2023mtbench}, mindful of LLM-judge inconsistency \citep{stureborg2024inconsistent}, bias against minority responses \citep{wang2023notfair}, and self-preference \citep{panickssery2024selffavor,bavaresco2024llmjudges}; the multi-vendor matrix is used only to characterise mapping uncertainty, not to validate occupation-coding correctness. Classical probability-sampling and discrete multivariate analysis \citep{cochran1977sampling,bishop1975discrete}, synthetic micro-data theory \citep{drechsler2011synthetic}, ISCO-08 \citep{ilo2012isco}, and ML/rule-based occupation coding \citep{schierholz2021occupation} ground the marginal-vs-joint framing.

\section{Validation Target: NPK-Korea}
\label{sec:resource}

NPK-Korea consists of one million records of adults aged $19$--$99$. There are $26$ columns in total: $1$ \texttt{uuid}, $7$ narrative fields, $6$ structured attribute fields, and $12$ demographic fields. Notable cardinalities are \texttt{family\_type} with 39 classes, \texttt{district} with 252 classes, and free-text \texttt{occupation} between $2$ and $40$ characters. The dataset card describes generation based on a proprietary probabilistic graphical model, NeMo Data Designer, and \texttt{google/gemma-4-31B-it}, and states the load-bearing assumption for this audit explicitly\footnote{NVIDIA Nemotron-Personas-Korea dataset card, \url{https://huggingface.co/datasets/nvidia/Nemotron-Personas-Korea}, accessed 2026-05-01.}:
\begin{quote}
``Demographic factors (sex, income, education, and major field) influence occupation assignment independently and interaction effects are not modelled.''
\end{quote}
The IAF audit of §\ref{sec:structural} contrasts this disclosed independence assumption against the corresponding joint structures in public Korean references \citep{nvidia2026npk}.

\subsection{Card-cited sources and audit feasibility}
The card lists five Korean public-data sources as sampling priors: KOSIS \citep{kosis2026openapi}, Supreme Court of Korea name statistics \citep{scourt2014family,efamily2026stats}, NHIS, KREI Food Consumption Behaviour Survey, and NAVER Cloud private seed data. KOSIS is the primary comparator and the Supreme Court name statistics anchor the name-cohort negative control; KEIS GOMS is added as a non-card public comparator for the major$\times$occupation joint. NHIS and KREI are deferred for microdata-access reasons and the NAVER Cloud seed is private and non-auditable. KoBBQ and LoRA downstream signals are auxiliary boundary checks in the released artefacts, not at the same evidence tier.

\section{Official-Reference Audit Protocol}
\label{sec:audit-framework}

The audit is built on three principles: central claims are restricted to experiments reproducible from public references alone, card-documented independence assumptions are placed at the centre of measurement, and negative controls are pre-specified from card-cited sampling sources (Supreme Court name statistics, KOSIS marital register) so that any PASS verdict is interpretable as design verification rather than post-hoc absence of joint structure.

\subsection{IAF primitive}
The Independence-Assumption Footprint (IAF) contrasts attribute combinations the dataset card describes as treated independently with the corresponding official or institutional joint structure. IAF is an instrument-defined operationalisation \citep{cronbach1955construct}: structural fidelity is operationalised as the KOSIS-vs-NPK Cramér's $V$ ratio plus category-level share diagnostics on card-disclosed factors. The direction of any sex$\times$occupation deviation is a fidelity signal, not a fairness verdict, because anti-stereotypical generation and joint-distribution infidelity are observationally indistinguishable from marginals alone. The audit anchors (KOSIS, KEIS, the Supreme Court name statistics, the MMA yearbook, KOSIS \texttt{DT\_1MR2060}) are themselves state classification products; IAF measures fidelity to those references rather than their underlying validity.

For a disclosed independence factor pair $(A,B)$ with a direct public joint table, we compute df-aware Cramér's $V$ \citep{bergsma2013cramer} on both the NPK and reference contingency tables and take the ratio
\begin{equation}
R_V(A,B) = V_{\mathrm{NPK}}(A,B)/V_{\mathrm{ref}}(A,B),
\quad \text{FLAG} \iff R_V<0.7.
\label{eq:leak}
\end{equation}
We additionally report a Feldman-style 4/5 parity band $r_{\mathrm{share}}\in[0.80,1.25]$ \citep{feldman2015disparate}, equivalent to $w\in[0.444,0.556]$ for a binary sex share, and a descriptive reference ratio $q_b=w_{\mathrm{NPK},b}/w_{\mathrm{ref},b}$ per occupation bucket. The three quantities are distinct: only $R_V$ enters the strict screening verdict, $r_{\mathrm{share}}$ is a within-NPK parity flag, and $q_b$ is a descriptive NPK-vs-reference contrast. The $0.7$ threshold is a fixed heuristic, not a calibrated equivalence to Feldman 4/5 and not a population significance test; threshold-perturbation sensitivity is reported in Table~\ref{tab:iaf_thresholds}. When the public source supplies a rate rather than a direct contingency table (military, name axes), $V_{\mathrm{ref}}$ is rule-implied from a synthetic reference table consistent with the institutional rule, and the resulting row is read as an institutional consistency screen rather than an association-retention estimate. A Pearl d-separation interpretation \citep{pearl2009causality} of card factorisation vs empirical reference is left to separate work.

\emph{Transferability scope.} Strict same-procedure transfer requires the target locale to have an occupation-level KSCO-equivalent classification and an occupation$\times$sex reference (NPK-Korea, NPK-USA). Adapted transfer changes the measurement grain (NPK-Japan industry-level). Cross-locale claims of §\ref{sec:cross-locale} establish strict transfer and only directionally confirm the adapted form.

For NPK-Korea, the key documented assumption is the sex/income/education/major independence in occupation assignment. We therefore select six externally anchored axes spanning occupation-linked joints, institutional life-course checks, and two negative controls; each axis consists of (i) the joint generated by NPK, (ii) a public reference, and (iii) a specified screening result or diagnostic deviation. The protocol map and numerical screening summary appear in Table~\ref{tab:iaf_thresholds}.

\begin{table}[!ht]
\centering\scriptsize
\setlength{\tabcolsep}{2pt}
\begin{tabular*}{\textwidth}{@{\extracolsep{\fill}}p{0.20\linewidth}p{0.17\linewidth}rrp{0.30\linewidth}p{0.12\linewidth}@{}}
\toprule
Axis & Reference & $V_{\mathrm{NPK}}$ & $V_{\mathrm{ref}}$ & Headline diagnostic & Screen / reading \\
\midrule
sex$\times$KSCO occupation & KOSIS LFP & $0.209$ & $0.362$\,(d) & male-dominated buckets $2.1$--$3.0\times$ KOSIS; $20.69$\,pp gap & \textbf{prod.-rule screen}$^\dagger$ \\
sex$\times$age$\times$LFP & KOSIS panel & $0.420$ & $0.386$\,(d) & young males $+16$\,pp; $60+$ $-13$\,pp & PASS + diag. \\
sex$\times$age$\times$military & MMA / conscription & $0.080$ & $0.224$\,(r) & flat $1.1$--$1.7\%$ NPK vs $26$--$29\%$ baseline & \textbf{inst.\ fail} \\
birth-year$\times$sex$\times$name & Supreme Court name prior & $0.205$ & $0.183$\,(r) & modern/vintage cohort structure reproduced & PASS (neg.\ ctrl.) \\
major$\times$occupation & KEIS GOMS & $0.080$ & $0.236$\,(d) & eng--hum $8$\_Plant gap $1.00$\,pp NPK vs $14.84$\,pp ref & \textbf{flagged} \\
sex$\times$marital & KOSIS \texttt{DT\_1MR2060} & $0.192$ & $0.168$\,(d) & max $|\Delta\,\mathrm{pp}|=2.88$ & PASS (neg.\ ctrl.) \\
\botrule
\end{tabular*}
\caption{Six structural-audit axes, references, $V$ values, and screen verdicts (rule in Eq.~\eqref{eq:leak}). $V_{\mathrm{ref}}$ source: (d) direct table, (r) rule-implied. $^\dagger$Production-rule mapping; reviewed-label variants cross the threshold (Layer 4).}
\label{tab:iaf_thresholds}
\end{table}
\FloatBarrier

\subsection{Metrics and reproducibility}
For age marginals we use KL divergence and 1-D Wasserstein distance, for sex marginals total variation, and for province Pearson $\chi^2$ and standardised residuals. For joint fidelity we use KL and Cramér's V. The KSCO crosswalk is a rule-based regex mapping with $96.1\%$ coverage on the 1M-row NPK. All non-API computation is run with Python 3.11, the \texttt{uv} dependency manager with a committed \texttt{uv.lock}, and seed 42 for stochastic steps only (deterministic table computations are seed-independent). The cumulative external-API spend across the audit is approximately USD~\$14, dominated by the three-LLM consensus matrix at under USD~\$5; cached API responses are released so that downstream replication does not re-incur this cost. KOSIS tables were downloaded via OpenAPI on 2026-05-01. The release accompanying this paper includes the KSCO-7 crosswalk and consistency rules, reference manifests for KOSIS/KEIS/MMA/Supreme Court, per-axis derived metric tables, the three-LLM consensus cache ($n=757$), permutation and bootstrap replicates, cross-locale audit caches for all seven NPK locales, and figure-generation scripts, under MIT licence at the anonymous mirror cited in the Declarations; upstream datasets retain their original licences and are not redistributed.

\subsection{Official-reference and annotation-light validation strategy}
\label{sec:annotation-free}

This audit recruits no external annotators and collects no new participant annotation. A non-blind first-author KSCO lookup review on boundary cases provides an auxiliary mapping-sensitivity check, not a new annotation benchmark. The design follows the LRE precedent of \citet{rigoutsterryn2020term}, who report that ``the magnitude of this task did not allow us to use multiple annotators'' and adopt consistency-over-multi-annotator-with-low-IAA as the working strategy. \citet{shardlow2022complex} criticises single-annotator datasets in LRE, so we anchor the quantitative claims to \emph{previously published} institutional and survey references (KSIC$\leftrightarrow$KSCO concordance from Statistics Korea, KEIS GOMS major$\times$occupation joint \citep{keis2020goms}, KoBBQ survey aggregate over $1{,}600$ Korean respondents \citep{jin2023kobbq}) rather than to first-author labels. Mapping uncertainty is probed via the multi-LLM consensus matrix and perturbation-sensitivity analysis. A blind multi-annotator extension is reserved as follow-up work, in line with \citet{song2026kolla}.

\section{Marginal Fidelity Baseline}
\label{sec:marginal-baseline}
\label{sec:fidelity}

We first quantify how closely NPK's core demographic spine matches KOSIS (Table~\ref{tab:marginal_baseline}). The aim of this section is to set the range over which the vendor's marginal-alignment claim holds before critiquing NPK. NPK is restricted to one million adult records and is contrasted with KOSIS \texttt{DT\_1B040M1\_1} (2025).

\begin{table}[tbp]
\centering\footnotesize
\setlength{\tabcolsep}{2pt}
\begin{tabular*}{\textwidth}{@{\extracolsep{\fill}}p{0.18\linewidth}p{0.27\linewidth}p{0.44\linewidth}@{}}
\toprule
Dimension & Metric / result & Reading \\
\midrule
Age & KL $=0.0037$ nats & strong marginal alignment \\
Sex & TV $=0.0006$ & strong marginal alignment \\
Province & mean $|\Delta_p|=0.06$\,pp & small but systematic residuals \\
Education & bachelors $+7.1$\,pp; graduate $+2.4$\,pp & measured over-representation \\
Marital status & unmarried $-9.3$\,pp & measured under-representation \\
Province$\times$sex$\times$age & KL $=0.0056$ nats & small global divergence, non-monotone old-age cells \\
\botrule
\end{tabular*}
\caption{Marginal and low-order fidelity baseline. Age/sex/province
align strongly with KOSIS, while education/marital status and some
high-age joint cells exhibit measurable deviations.}
\label{tab:marginal_baseline}
\end{table}

The province residuals are small in magnitude but systematic. Metropolitan cities tend to be over-represented and provinces (the ``do''-level administrative units) under-represented. In the province$\times$sex$\times$age joint with $2{,}754$ cells, every province exhibits the same pattern of over-represented 99-year-old males together with empty cells for some $90$--$97$-year-old males. With $n=1$\,M the $\chi^2$ test is power-saturated and every province carries $|z|>2$, so the load-bearing practical-equivalence summary is the mean $|\Delta_p|=0.06$\,pp rather than the per-province rejection. Marginal alignment is therefore strong, but cell-level artefacts emerge as soon as downstream users start conditioning on slices.

This baseline opens the next question: while NPK matches the age/sex/province marginals reasonably well, does it also preserve the joint distributions that matter for occupations and the life course in Korea?

\section{Structural Fidelity Audit: IAF Axes}
\label{sec:structural}

This section contains the central results of the paper. The NPK card explicitly disclaims joint modelling. In its own words, demographic factors (sex, income, education, major field) influence occupation assignment independently and interaction effects are not modelled \citep{nvidia2026npk}. We contrast the empirical footprint of this documented assumption against externally anchored Korean reference groups plus an additional marital-status control.

\subsection{Sex-by-KSCO occupation axis}
\label{sec:fair-pillar-ksco}

\begin{figure}[tbp]
\centering
\includegraphics[width=\linewidth]{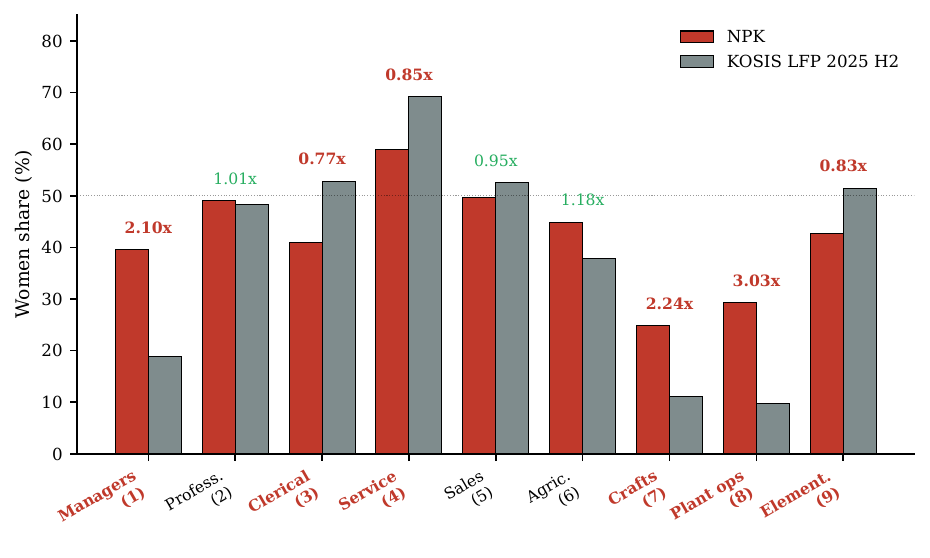}
\caption{Female share by KSCO major class: NPK vs KOSIS.}
\label{fig:pillar1}
\end{figure}

The $2{,}120$ free-text NPK occupations are mapped to the 9 KSCO major classes plus a \texttt{0\_무직} (``not in employment'') bucket. Coverage is $96.1\%$. The reference is KOSIS \texttt{DT\_1ES3B32J} (national occupation/sex employed persons, 2025\,H2). Following the official-reference strategy described in §\ref{sec:annotation-free}, we stress-test KSCO mapping uncertainty across four diagnostic layers: round-trip closure on the 9 official class labels (Layer 1), random label-noise stability of the production rule (Layer 2), a multi-vendor LLM panel-of-judges consensus matrix (Layer 3), and a non-blind first-author KSCO-7 lookup review on $n=757$ stratified labels (Layer 4).

\textbf{Layers 1--2.} Round-trip closure on $35$ official KSCO class names routes $25/35$ correctly; misses (UNMAPPED 종사자 suffixes, 기사 misrouted to 2\_전문가, 경비원/청소원 misrouted to 4\_서비스 instead of KSCO-7 9\_단순노무) seed Layer 4 boundary cases. A random label-noise stress test (flipping $5$/$10$/$20\%$ of NPK$\to$KSCO assignments, 100 replicates) keeps the production-rule screen below threshold in $100/100$ iterations even at $20\%$ noise.

\textbf{Layer 3.} Following the panel-of-judges design of \citet{verga2024juries}, we classify $757$ stratified NPK labels with Claude \texttt{claude-opus-4-5}, Gemini \texttt{gemini-2.5-flash}, and HyperCLOVA-X \texttt{HCX-005} at temperature $0$, building a 4-way agreement matrix with the rule mapper. The three judges share public KSCO institutional knowledge and Gemini shares Google's pretraining lineage with the \texttt{gemma-4-31B-it} backbone, so the rule-vs-LLM band ($\kappa=0.520$--$0.561$) is the conservative anchor; Claude--Gemini agreement is interpreted with a self-preference caveat \citep{panickssery2024selffavor}. Pairwise system-agreement $\kappa$ is $0.520$--$0.561$ rule-vs-any-LLM and $0.839$--$0.934$ LLM-vs-LLM; $403/757$ ($53.2\%$) labels reach 4-way unanimity and $685/757$ ($90.5\%$) reach $\geq 3/4$ majority. The 4-rater omnibus Fleiss $\kappa$ \citep{fleiss1971kappa} on the $11$ categories is $0.705$, reported as a panel-of-judges descriptive statistic \citep{verga2024juries}; the \citet{landis1977kappa} ``substantial'' label does not transfer because the four raters are not independent human annotators. The matrix characterises automated-coder disagreement but does not validate occupation-coding accuracy; the public references (KSIC$\leftrightarrow$KSCO concordance from Statistics Korea, KEIS GOMS \citep{keis2020goms}) anchor the comparison target rather than every NPK mapping decision \citep{shardlow2022complex}.

\textbf{Layer 4 (auxiliary).} The first author (native Korean speaker, eight years of AI research experience) reviewed the $757$ stratified labels against KSCO-7 4-digit codes; $686/757$ ($90.6\%$) received explicit codes, with the rest falling back to major class. Categories are KSCO 9-major + \texttt{0\_무직} + \texttt{A\_군인} (KSCO-7 major group 0, excluded from the KOSIS LFP frame; $5{,}391$ records) + \texttt{UNMAPPED} (0 records). Six boundary cases (§\ref{sec:layer4-rules}) override automated-coder consensus by citing KSCO-7. Prefill columns were visible, so the descriptive agreement rates ($75.4\%$ rule, $76.0$--$76.1\%$ Claude/Gemini, $67.6\%$ HyperCLOVA-X) are not blind IAA and the reviewed-vs-rule overlap is a parity-flag sensitivity probe rather than independent corroboration. The reviewed lookup affects $30{,}737$ NPK records ($3.07\%$ of 1M; $3.33\%$ of the reviewed covered set $923{,}225$).

Applying both mappings to NPK 1M (age $\geq 19$), $6/9$ KSCO buckets trigger the within-NPK parity flag under either mapping with $0$ parity-flag-changing buckets (Table~\ref{tab:layer4_sensitivity}). The reviewed-label-only view gives $R_V=0.725$ and the same-population reviewed+rule-fallback view gives $R_V=0.713$; both cross the fixed $R_V<0.7$ threshold while remaining below full reference retention. A seed-42 sex-label permutation null preserving the NPK KSCO marginal ($1{,}000$ replicates) places the observed production-rule $R_V=0.578$ far above the independence-generating null (upper-tail empirical $p=0.001$, plus-one estimate). The directional attenuation conclusion of §\ref{sec:fair-pillar-ksco} therefore holds across mappings, but the strict screening verdict is mapping-sensitive. Because both mappings share the same NPK records and KOSIS reference, genuinely independent verdict tests come from KEIS GOMS (§\ref{sec:fair-pillar-major}) and KOSIS \texttt{DT\_1MR2060} (§\ref{sec:fair-pillar-marital}).

\emph{Robustness summary.} The strict-screen verdict is stable across $\tau\in[0.60, 0.70]$; the male-dominated-bucket ratio survives direct age standardisation against the 2020 Korean Census KSCO age structure (DT\_1PC2010); and $4/9$ buckets remain outside the q-band under a Kish DEFF=$1.5$ adjustment for KOSIS LFS sampling uncertainty. The full threshold grid, age-standardisation per bucket, and DEFF=$1.5$/$2.0$ robust flag tables are released as derived metric artefacts.

\begin{table}[tbp]
\centering\scriptsize
\setlength{\tabcolsep}{1.8pt}
\begin{tabular}{@{}lrrrrrrrll@{}}
\toprule
KSCO bucket & $n_r$ & $n_v$ & $w_r$ & $w_v$ & $w_K$ & $\Delta_r$ & $\Delta_v$ & parity $r$ & parity $v$ \\
\midrule
1\_관리자 & 16{,}372 & 18{,}419 & 0.395 & 0.385 & 0.188 & $+0.207$ & $+0.196$ & out & out \\
2\_전문가 & 132{,}465 & 92{,}097 & 0.490 & 0.473 & 0.483 & $+0.007$ & $-0.010$ & in & in \\
3\_사무 & 122{,}685 & 133{,}772 & 0.409 & 0.420 & 0.528 & $-0.119$ & $-0.108$ & out & out \\
4\_서비스 & 113{,}402 & 67{,}404 & 0.590 & 0.752 & 0.691 & $-0.101$ & $\mathbf{+0.060}$ & out & out \\
5\_판매 & 43{,}054 & 43{,}006 & 0.496 & 0.496 & 0.525 & $-0.029$ & $-0.029$ & in & in \\
6\_농림어업 & 2{,}192 & 104 & 0.448 & 0.548 & 0.379 & $+0.070$ & $+0.169$ & in & in \\
7\_기능원 & 45{,}556 & 42{,}273 & 0.249 & 0.286 & 0.111 & $+0.138$ & $+0.175$ & out & out \\
8\_장치조작 & 71{,}283 & 69{,}186 & 0.293 & 0.292 & 0.097 & $+0.196$ & $+0.195$ & out & out \\
9\_단순노무 & 47{,}024 & 84{,}224 & 0.427 & 0.444 & 0.514 & $-0.087$ & $-0.071$ & out & out \\
\midrule
\multicolumn{9}{@{}l}{parity-flag-changing buckets:} & $\mathbf{0/9}$ \\
\multicolumn{9}{@{}l}{$\Delta$ sign agreement (rule vs review):} & $\mathbf{8/9}$ \\
\botrule
\end{tabular}
\caption{Crosswalk-choice sensitivity: rule-based ($r$) vs reviewed exact-string KSCO lookup ($v$, $n=757$). $w_K$ is the KOSIS reference share.}
\label{tab:layer4_sensitivity}
\end{table}
\FloatBarrier

Four of nine buckets have $95\%$ reference-ratio bands entirely outside $[0.8, 1.25]$ and remain outside under Bonferroni across the 9-bucket family; with $n \gtrsim 44{,}000$ per bucket the bands measure within-NPK resampling, not generative-model variance. The within-NPK parity flag and the reference-ratio diagnostic are distinct: $4$\_서비스 ($q\approx 0.85$) and $9$\_단순노무 ($q\approx 0.83$) sit inside the reference band yet remain parity-flagged because $w_{\mathrm{NPK}}$ is far from $0.5$. Rule/LLM/reviewer overlaps reflect convergent lookup against shared KSCO-7 documentation rather than independent judgments.

NPK exhibits a flatter gender distribution than the Korean LFP (Fig.~\ref{fig:pillar1}). In the three most male-dominated KSCO buckets (Plant/Machine, Crafts, Managers), the female share in NPK is $2.1$--$3.0\times$ that of KOSIS. In the three most female-dominated buckets (Service, Clerical, Elementary), it is $8$--$12$\,pp lower than KOSIS. The audit-relevant signal is the conjunction of low sex$\times$KSCO Cramér's $V$, large absolute gaps, and the descriptive NPK-to-KOSIS reference ratios; per-bucket cell values are released as a derived metric artefact.

\subsection{KSCO-7 mapping consistency rules}
\label{sec:layer4-rules}

The Layer 4 review applies 14 decision rules to resolve disputed and boundary cases when mapping NPK free-text occupations into KSCO-7 (Table~\ref{tab:rules_summary}). Each rule cites its KSCO-7 4-digit code where unambiguous and states how it resolves prefilled LLM consensus or prefix-matching rule disagreements. The lookup is propagated by exact raw \texttt{occupation}-string join from the released reviewed-label artefact to $923{,}225$ adult NPK records ($550{,}485$ in the 9 comparable buckets, $367{,}349$ in \texttt{0\_무직}, $5{,}391$ in \texttt{A\_군인}, $76{,}775$ \texttt{UNMAPPED}); no fuzzy matching, semantic expansion, or unreviewed-label propagation is used. Two rules need rationale beyond the table. Rule 12 (``전직 X, 현재 구직중'' $\to$ base occupation X) preserves NPK base-occupation identity for the sex-LFP analysis (treating job-seekers as \texttt{0\_무직} would erase ${\sim}74$K records, ${\sim}7.4\%$ of adults; 전직 군인 maps to \texttt{A\_군인}). Rule 13 (Military $\to$ \texttt{A\_군인}) aligns with the KOSIS LFP convention; a Rule 13 ablation leaves the strict screen unchanged (literal \texttt{A\_군인} exclusion gives $R_V=0.612$; the default free-text mapper sends all 17 military labels to 4\_서비스 and reproduces the headline $R_V=0.578$; all four variants retain $6/9$ parity flags and remain below the $R_V<0.7$ screen). Rule 14 (boundary cases) is detailed in the released artefact.

\begin{table}[tbp]
\centering\footnotesize
\setlength{\tabcolsep}{3pt}
\begin{tabular*}{\textwidth}{@{\extracolsep{\fill}}lp{0.34\linewidth}p{0.18\linewidth}p{0.18\linewidth}p{0.16\linewidth}@{}}
\toprule
\# & Rule (Korean pattern, condensed) & Target KSCO-7 & Override of & NPK impact \\
\midrule
1 & 일반 경비원 (건물/시설/아파트) & 9\_단순노무 (9421) & LLM 4-way 4\_서비스 & 28{,}754 \\
2 & 청소원, 환경미화원 & 9\_단순노무 (9512/9519/9521) & rule & (n/a) \\
3 & 안내원/도우미/해설사/점검원 등 & 4\_서비스 (432X) & rule 3\_사무 & 6{,}627 \\
4 & 텔레마케터 (vs 전화상담원) & 5\_판매 (5314 outbound) & rule 3\_사무 & 1{,}827 \\
5 & 우편집배원 & 4\_서비스 (4222) & rule 8\_장치조작 & 1{,}915 \\
6 & X 운전원, 택배원, 배달원 & 8\_장치조작 (86XX) & rule & 29{,}710 \\
7 & X 기계 조작원/조립원 & 8\_장치조작 & rule 7\_기능원 & 29{,}538 \\
8 & 설치원/수리원/통신기술자 & 7\_기능원 (74XX--77XX) & rule & 13{,}779 \\
9 & 시험원/연구원 (어업 포함) & 2\_전문가 (21XX/22XX) & LLM 6\_농림어업 (어업 시험원) & 16{,}349 \\
10 & 제도사/디자이너 & 2\_전문가 (2351/285X) & rule & 3{,}102 \\
11 & 건설 기술자 vs 기능공 & 2\_전문가 (2143) / 7\_기능원 & rule & 2{,}886 \\
12 & ``전직 X, 현재 구직중'' & base occupation X & rule 0\_무직 & 12{,}626 \\
13 & 군인 (현역+직업+전직) & A\_군인 (KSCO-7 group 0) & rule (varies) & 5{,}391 \\
14 & Boundary cases (released artefact) & (per-row) & LLM 4-way / rule & 7{,}505 \\
\botrule
\end{tabular*}
\caption{The 14 KSCO-7 consistency rules applied in the Layer 4 reviewed lookup.}
\label{tab:rules_summary}
\end{table}

Six boundary cases (e.g., 건물/시설/아파트 경비원, 어업 시험원, 가사 도우미, 보육교사) were resolved by citing the KSCO-7 4-digit code over automated-coder consensus; the resolved-vs-consensus comparison is in the released artefacts. Explicit KSCO-7 4-digit codes are assigned to $686/757$ ($90.6\%$) labels; the remaining $71$ fall back to major-class only. When these 14 rules are joined back to NPK 1M by exact occupation string, the 9-class parity-flag pattern is identical to the rule-based regex mapper at the bucket level (Table~\ref{tab:layer4_sensitivity}); $0/9$ parity-flag-changing buckets. A leave-one-rule-out sweep shows that $9/14$ rule removals preserve the baseline $6/9$ parity-flag count and $5/14$ shift it by $\pm 1$ (range $[5/9, 7/9]$); in every single-rule removal at least $5/9$ buckets are still parity-flagged. No single rule is load-bearing for the qualitative parity-flag or flattening conclusion. Per-rule LOO counts are released as a derived metric artefact.

\subsection{Sex-by-age-by-LFP axis}
\label{sec:fair-pillar-mcurve}

KOSIS \texttt{DT\_1ES3B03J} and \texttt{DT\_1ES3B14J} (2025\,H2) provide LFP rates by sex and age band. The NPK analogue is $1-\Pr(\texttt{occupation}\in\text{not-in-LF})$ via the KSCO crosswalk (not-in-LF keyword set: \texttt{무직/실업/구직/주부/전업주부/학생/대학생/고등학생/중학생/초등학생/유치원/퇴직자/은퇴/은퇴자/자영업 폐업/실업자}); treating \texttt{구직} as not-in-LF is opposite to KOSIS ILO standard, so part of the ages-19--29 deviation is a definitional gap. Under the $V$-ratio rule the axis passes ($V_{\mathrm{NPK}}=0.42$ vs $V_{\mathrm{ref}}=0.39$), but the band-wise rates show directionally asymmetric deviations that a directionless Cramér's $V$ can mask; we therefore report the deviation as a diagnostic rather than as an association-retention flag.

\begin{table}[tbp]
\centering\footnotesize
\setlength{\tabcolsep}{3.5pt}
\begin{tabular*}{\textwidth}{@{\extracolsep{\fill}}lrrrrrr@{}}
\toprule
band & K-M & N-M & $\Delta$pp & K-F & N-F & $\Delta$pp \\
\midrule
19--29 & .452 & .612 & \textbf{+15.97} & .490 & .449 & $-4.13$ \\
30--39 & .895 & .841 & $-5.43$ & .757 & .721 & $-3.61$ \\
40--49 & .922 & .845 & $-7.72$ & .708 & .731 & $+2.30$ \\
50--59 & .887 & .828 & $-5.88$ & .694 & .702 & $+0.81$ \\
60+ & .578 & .442 & \textbf{$-13.61$} & .410 & .281 & \textbf{$-12.90$} \\
\botrule
\end{tabular*}
\caption{LFP rate by sex$\times$age band: KOSIS (K) vs NPK (N).}
\label{tab:lfp_mcurve_full}
\end{table}

NPK male LFP at ages $19$--$29$ is $+15.97$\,pp higher than KOSIS (Table~\ref{tab:lfp_mcurve_full}). Conversely, both males and females aged $60+$ are about $-13$\,pp lower than KOSIS. This axis is therefore not an association-retention flag, but it is a diagnostically important deviation for silicon-sample users conditioning on labour-market slices.

\subsection{Sex-by-age-by-military axis}
\label{sec:fair-pillar-military}

\begin{figure}[tbp]
\centering
\includegraphics[width=\linewidth]{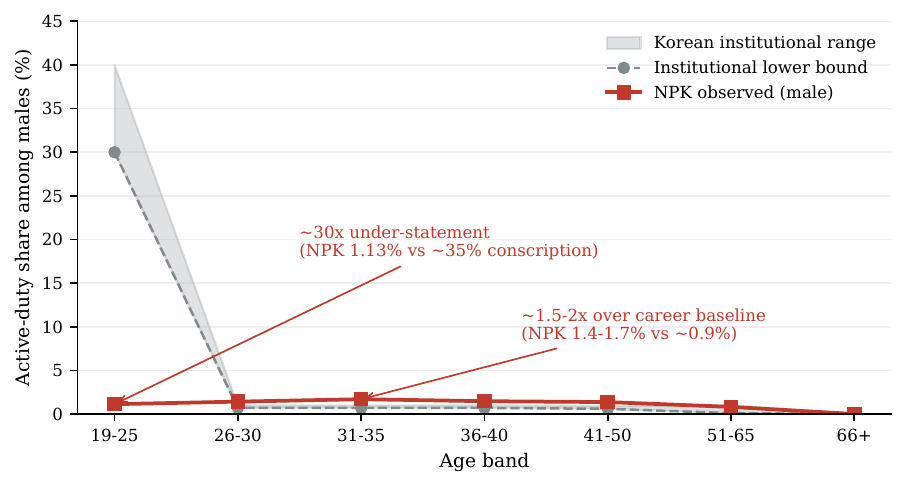}
\caption{NPK male active-duty share by age band.}
\label{fig:pillar3}
\end{figure}

\texttt{military\_status} is binary (active / non-active). Korean males aged $19$--$22$ are subject to mandatory active-duty service. Using KOSIS \texttt{DT\_1B040M1\_1} 2024 male resident-registration counts for ages $19$--$22$ ($1.010$M) and MMA 2024 yearbook age-specific entries \citep{mma2024yearbook}, a stock approximation gives a $26$--$29\%$ active-duty rate for ages $19$--$22$ ($26.1\%$ for 현역 only; $28.6\%$ including 사회복무요원/보충역 at $21$ months). After discharge personnel transition to reservist status. Career military (officers, warrants, NCOs; ${\sim}210$K against ${\sim}21$M adult males) sustains a baseline of $0.7$--$1.0\%$ pre-retirement and ${<}0.2\%$ after age $55$--$60$. Female active-duty share is volunteer-only and ${\leq}1\%$ at all ages. This axis is closer to an institutional baseline check than to a direct contingency table.

\begin{table}[tbp]
\centering\footnotesize
\begin{tabular*}{\textwidth}{@{\extracolsep{\fill}}lrrl@{}}
\toprule
band & male act.-duty \% & female act.-duty \% & vs. institutional baseline \\
\midrule
19--25 & 1.13 & 0.11 & far below mandatory $26$--$29\%$$^\dagger$ \\
26--30 & 1.42 & 0.13 & ${\sim}1.5\times$ over career baseline ${\sim}0.9\%$ \\
31--35 & 1.70 & 0.18 & ${\sim}1.9\times$ over career baseline ${\sim}0.9\%$ \\
36--40 & 1.47 & 0.19 & ${\sim}1.6\times$ over career baseline ${\sim}0.9\%$ \\
41--50 & 1.37 & 0.17 & ${\sim}1.7\times$ over career baseline ${\sim}0.8\%$ \\
51--65 & 0.82 & 0.10 & mixed pre-/post-retirement baseline \\
66+ & 0.00 & 0.00 & natural at $0$ post-retirement (age $55$--$60$) \\
\botrule
\end{tabular*}
\caption{NPK \texttt{military\_status}=active share by age band vs Korean institutional baseline. $^\dagger$NPK schema is binary; comparison to the conscription tier is qualitative.}
\label{tab:military_full}
\end{table}

NPK shows male active-duty roughly flat at $1.13$--$1.70\%$ across ages $19$--$50$ (Fig.~\ref{fig:pillar3}, Table~\ref{tab:military_full}), with the share at $31$--$35$ higher than at $19$--$25$, the reverse of the Korean conscription pattern. Without generation logs or a direct public contingency table we treat this as an institutional consistency failure rather than evidence about the exact sampling mechanism (§\ref{sec:limit-attribution}).

\emph{Schema gap caveat.} The binary \texttt{military\_status} collapses Korea's six-tier structure (현역, 보충역, 예비역, 병역특례, 면제, 미필); ${\sim}60\%$ of Korean males in their thirties are reservists (예비역) and are lumped under ``non-active duty.'' A precise age-conditional audit would require an NPK schema extension.

\subsection{Birth-year-by-sex-by-name axis}
\label{sec:fair-pillar-names}

The card cites the Supreme Court of Korea (S2) birth-year$\times$sex$\times$name as a sampling source. NPK first names are extracted from persona narratives (regex \texttt{[가-힣]\{2,5\}\,씨[는가께도이]}, $96.7\%$ extraction) and compared against the Korean naming-fashion cohort prior as a negative control. NPK reproduces the cohort structure: vintage-name share decreases by $56\times$ from pre-1950 (top names 영자/정자/순자/춘자, $13.95\%$ vintage) to 1990--2007 (top names 유진/지원/민지/수빈, $5.24\%$ modern surge, $0.25\%$ vintage), and the surname distribution matches Korea Census 2015 within $\pm 0.5$\,pp on the top 4. This indicates that NPK's narrative samples names jointly with age, in contrast with the occupation and military independence patterns.

\subsection{Bachelors-field-by-occupation axis}
\label{sec:fair-pillar-major}

The card lists major as one of the independence factors in occupation assignment. The external reference is KEIS GOMS (public microdata; current-job comparator, $n=11{,}823$ weighted) mapped to 9 KSCO major classes; the GOMS cohort is 4-year university graduates and not a super-set of NPK \texttt{bachelors\_field} records, so part of the $\Delta$pp gap is attributable to universe differences rather than generative misalignment, and we interpret the comparison as conditional mismatch within the KEIS graduate universe. A universe-matched sensitivity restricting NPK to ages $23$--$26$ with \texttt{education\_level}=\texttt{4년제 대학교} ($9{,}589$ records) does not attenuate the headline; Engineering$\to$8\_Plant moves from $-14.33$ to $-14.45$\,pp and Education/ICT/Health$\to$2\_Profess gaps widen further. A leave-one-major-out sweep keeps the Engineering$\to$8\_Plant gap in $[-14.45, -10.61]$\,pp across $19$ leave-outs (max shift $3.84$\,pp on dropping $3$\_Clerical), so the headline is not driven by any single major or field.

\begin{table}[tbp]
\centering\footnotesize
\setlength{\tabcolsep}{4pt}
\begin{tabular}{@{}lrrrr@{}}
\toprule
KEIS GOMS comparison & KEIS grads & NPK & $\Delta$pp & $|\Delta|>5$? \\
\midrule
Engineering/Manufacturing/Construction $\to$ 8\_Plant & 18.13 & 3.80 & $-14.33$ & yes \\
Engineering/Manufacturing/Construction $\to$ 2\_Profess & 53.05 & 29.7 & $-23.35$ & yes \\
Engineering/Manufacturing/Construction $\to$ 3\_Clerical & 19.09 & 44.3 & $+25.21$ & yes \\
Agriculture/Fisheries/Veterinary $\to$ 6\_Agric & 9.19 & 0.40 & $-8.79$ & yes \\
Education $\to$ 2\_Profess & 82.96 & 40.2 & $-42.76$ & yes \\
Information/Communication Tech.\ $\to$ 2\_Profess & 65.28 & 25.9 & $-39.38$ & yes \\
Health/Welfare $\to$ 2\_Profess & 79.50 & 44.3 & $-35.20$ & yes \\
Service $\to$ 4\_Service & 34.04 & 5.20 & $-28.84$ & yes \\
Arts/Humanities $\to$ 8\_Plant & 3.29 & 2.80 & $-0.49$ & no \\
\botrule
\end{tabular}
\caption{$P(\text{KSCO}\mid\text{major})$: KEIS GOMS \citep{keis2020goms} 4-year graduate universe vs NPK degree-holders.}
\label{tab:major_keis_validated}
\end{table}

The major$\times$occupation joint shows a large KEIS-universe conditional mismatch (Table~\ref{tab:major_keis_validated}): engineering graduates appear in $8$\_Plant at $18.13\%$ vs $3.80\%$ in NPK; agriculture/veterinary graduates at $9.19\%$ in $6$\_Agric vs $0.40\%$; Education/ICT/Health/Welfare graduates in $2$\_Professionals carry $-35$ to $-42$\,pp gaps that a $5$--$10$\,pp cohort-aging adjustment cannot absorb. The KEIS GOMS reference is the $2018$ wave (release $2020$), a six-year vintage gap to NPK (released $2026$); we treat the reported $\Delta$ as the joint magnitude of cohort universe, vintage gap, and NPK generative attenuation, with vintage and cohort dominating only the smaller-magnitude rows. A KLIPS-23-wave triangulation is deferred (§\ref{sec:limit-keis-cohort}).

\subsection{Sex-by-marital-status axis (additional source)}
\label{sec:fair-pillar-marital}

As a second negative control we compare NPK marital status against KOSIS \texttt{DT\_1MR2060} (2024-12), aggregated to the national level after harmonising NPK \texttt{미혼/배우자있음/사별/이혼} with KOSIS \texttt{미혼/유배우/사별·이혼} and restricting to adults ($\geq 19$).

\begin{table}[tbp]
\centering\footnotesize
\setlength{\tabcolsep}{4pt}
\begin{tabular*}{\textwidth}{@{\extracolsep{\fill}}llrrrrr@{}}
\toprule
Sex & Marital & NPK \% & KOSIS \% & $\Delta$pp \\
\midrule
남자 & 미혼 & 30.85 & 32.87 & $-2.02$ \\
남자 & 유배우 & 60.49 & 58.30 & $+2.19$ \\
남자 & 사별·이혼 & 8.66 & 8.83 & $-0.17$ \\
여자 & 미혼 & 20.63 & 23.51 & $-2.88$ \\
여자 & 유배우 & 58.04 & 56.90 & $+1.15$ \\
여자 & 사별·이혼 & 21.32 & 19.59 & $+1.73$ \\
\botrule
\end{tabular*}
\caption{Sex $\times$ marital status: NPK vs KOSIS share and
$\Delta$.}
\label{tab:pillar6_marital}
\end{table}

The marital axis does not trigger the screening rule (Table~\ref{tab:pillar6_marital}); combined with the name cohort, it is the second pre-specified negative control. NPK aligns with KOSIS marginals on attributes the card does not flag as independence factors, while the strongest failures concentrate in occupation-linked joints. Both controls may partly reflect direct sampling from these references, so in the Lipsitch-strict sense \citep{lipsitch2010negativecontrols} they are design-verification probes rather than confounder-detecting controls; the name axis (non-KOSIS Supreme Court source) is the genuinely vendor-independent control, while the marital axis is a weaker within-KOSIS check. Structural negative controls (name$\times$occupation joint, sex$\times$marital$\times$age joint) are left to future work.

\subsection{Cross-locale IAF replication across the NPK family}
\label{sec:cross-locale}

\begin{figure}[tbp]
\centering
\includegraphics[width=\linewidth]{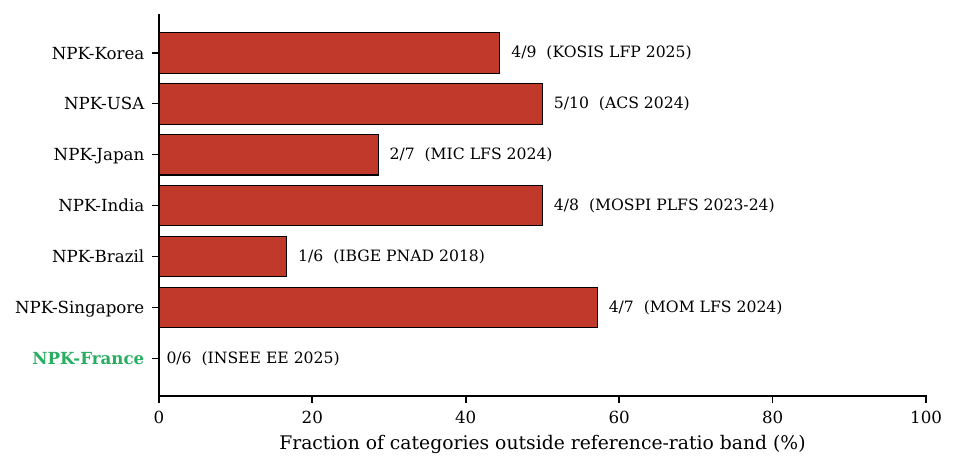}
\caption{Occupation-axis reference-ratio diagnostic flags under
locale-specific references across 7 NPK locales. Counts are
transferability diagnostics, not comparable effect sizes or a locale
ranking.}
\label{fig:cross_locale}
\end{figure}

\emph{Positionality.} The first author is a Korean researcher; this section presents IAF as a transferability demonstration rather than as an authoritative audit of any non-Korean locale. Local-expert audits are recommended future work. Flag counts depend on the available reference taxonomy (PCS-6 for France, SOC-10 for the USA, ISCO-8 for India) and on the LFS-vs-Census reference choice; they are not effect sizes and should not be read as a locale ranking.

To test whether the IAF primitive transfers beyond NPK-Korea, we apply the same reference-ratio diagnostic to all six other NPK country variants \citep{nvidiaPersonasCollection,nvidia2026npk,nvidiaPersonasUSA,nvidiaPersonasJapan,nvidiaPersonasIndia,nvidiaPersonasBrazil,nvidiaPersonasSingapore,nvidiaPersonasFrance}, each against its own official labour-force or census reference.

\textbf{NPK-USA against ACS 2024.} NPK-USA contains one million records with \texttt{country=USA}. The reference is the U.S.\ American Community Survey (ACS) 2024 1-year estimates, taken from the Data USA portal (2026-05-08 access), which sources Census Bureau ACS PUMS microdata \citep{acs2024datausa}. The ACS provides occupation $\times$ sex statistics broadly comparable to BLS CPS Annual Averages; the direct BLS retrieval below agrees within ${\pm}3$\,pp for the three detailed occupations whose LEU women and total series both resolved.

In NPK-USA, IAF occupation axis falls outside the reference-ratio band in $5/10$ occupations. The most extreme $\Delta$pp is $+46.5$\,pp ($12.53\times$) for construction laborers; drivers/sales workers/truck drivers (combined under one NPK label) are $+24$\,pp; software developers are $+18.5$\,pp; customer service representatives are $-26.0$\,pp, flattening the female-dominant pattern in the ACS. This is directionally consistent with the ``gender flattening'' pattern of NPK-Korea in §\ref{sec:fair-pillar-ksco} (over-feminising male-dominated occupations and under-feminising female-dominated ones), but we treat it as a transferability signal rather than a full locale-specific audit of NPK-USA. A direct BLS LEU-series query agrees with the ACS-via-Data-USA reference within ${\pm}3$\,pp on the occupations for which both series IDs resolved.

\textbf{NPK-Japan, industry-level cut.} NPK-Japan labels are JSIC-inspired free-text (industry $\times$ firm-size $\times$ career-stage; e.g., ``介護福祉業 中堅'' eldercare/welfare $\times$ mid-tier firm) rather than occupation taxonomy codes. We therefore adapt IAF to industry $\times$ sex against Japan's Labour Force Survey 2024 annual averages \citep{japanstathandbook2025}, at the LFS major-division level. The industry-adapted IAF falls outside the reference-ratio band in $2/7$ categories ($5/10$ NPK-USA, $4/9$ NPK-Korea); both flags show over-feminisation (Agriculture/Forestry $+14.6$\,pp, Construction $+5.2$\,pp). A male-dominated-category over-feminisation signal appears in all three comparisons; the smaller NPK-Japan magnitude indicates closer industry-level marginal alignment with LFS than the occupation-level alignment in Korea/USA. The IAF primitive therefore operationalises from occupation-level (Korea/USA) to industry-level (Japan) with a structural adaptation, but exact magnitudes should not be compared as if the grains were identical.

\emph{Caveats.} Segregation indices differ across industry$\times$sex and occupation$\times$sex grains \citep{cotter2003segregation}, so the smaller NPK-Japan deviation may be a function of grain rather than superior generative quality. A 林業 (forestry) sensitivity check shifts the NPK agriculture/forestry women \% by only $-0.05$\,pp, ruling out a NPK forestry undersampling artefact. Japanese agriculture sex composition is also shaped by family-business dynamics (spouse co-employment, ageing farming households), so the $+14.6$\,pp gap should not be read as a simple NPK generative-pattern artefact.

\textbf{IAF on 4 additional locales (India, Brazil, Singapore, France).} We also ran IAF occupation axis on India (MOSPI PLFS 2023--24 Usual Status PS+SS), Brazil (IBGE PNAD 2018), Singapore (MOM 2024, resident labour force excluding foreign domestic workers per MOM's published universe), and France (INSEE 2025) \citep{indiaPLFS2024,ibgePNAD2018,singaporeMOM2024,insee2025ee}.

\begin{table}[tbp]
\centering\footnotesize
\setlength{\tabcolsep}{3pt}
\begin{tabular*}{\textwidth}{@{\extracolsep{\fill}}lllrl@{}}
\toprule
NPK locale & LFS flags / extreme ($\Delta$pp) & Census flags & $\Delta$ flags & LFS source year \\
\midrule
NPK-Korea & 4/9, 8\_장치조작 ($+19.6$) & --- & --- & 2025\,H2 \\
NPK-USA & 5/10, construction\_laborer ($+46.5$) & --- & --- & 2024 \\
NPK-Japan & 2/7, 農業·林業 ($+14.6$) & 2/7 & $0$ & 2024 \\
NPK-India & 4/8, Service \& Sales Div 5 ($+44.1$) & 7/8 & \textbf{+3} & 2023--24 \\
NPK-Brazil & 1/6, ISCO 7 construction ($+6.2$) & 2/6 & $+1$ & 2018 \\
NPK-Singapore & 4/7, Service \& Sales ($-29.3$) & 4/7 & $0$ & 2024 \\
\textbf{NPK-France} & \textbf{0/6}, Artisans ($-5.9$) & 1/6 & $+1$ & 2025 \\
\botrule
\end{tabular*}
\caption{Occupation-axis reference-ratio flags across 7 NPK locales under LFS and (where applicable) Census re-anchor references.}
\label{tab:cross_locale_4countries}
\end{table}

\textbf{Locale-specific reference alignment.} Across the 7 NPK locales (Fig.~\ref{fig:cross_locale}, Table~\ref{tab:cross_locale_4countries}), NPK-Korea ($4/9$), NPK-USA ($5/10$), NPK-India ($4/8$), and NPK-Singapore ($4/7$) exhibit multiple flags with large magnitudes; NPK-Japan ($2/7$) shows a smaller magnitude partly because of its industry-level grain. NPK-Brazil ($1/6$) and NPK-France ($0/6$ out-of-band, $\Delta_{\max}=-5.9$\,pp) register fewer flags but at sharply reduced reference cardinality (8 unique NPK-France occupation labels, 6-group PCS reference), a known reference-ratio confounder; these flag counts are not locale quality rankings. A cardinality-control sensitivity collapsing high-cardinality Korea/USA/India audits to a six-role PCS-6-style grain (manager+professional, clerical, service+sales, agriculture, craft+trades, operators+elementary) attenuates Korea $4/9\to 2/6$ (max $|\Delta|=13.8$\,pp), USA $5/10\to 4/6$, and India $4/8\to 2/6$; taxonomy grain is a real count attenuator but Korea does not collapse to $0$--$1/6$, so France's $0/6$ should be read as grain-confounded rather than explained by grain alone.

\textbf{Census re-anchor.} The NPK locale cards \citep{nvidiaPersonasUSA,nvidiaPersonasJapan,nvidiaPersonasIndia,nvidiaPersonasBrazil,nvidiaPersonasSingapore,nvidiaPersonasFrance} cite each country's Census rather than LFS as the grounding source \citep{usCensusACS2024,japanCensus2020,indiaCensus2011,ibgeCensus2022,singaporeCensus2020,inseeRecensement2022}, so we re-ran the audit against the Census-cited reference for the 5 re-anchorable locales (right two columns of Table~\ref{tab:cross_locale_4countries}). Four are stable to within $\pm 1$ flag under the reference change; only NPK-India shifts substantially ($4\to 7$, consistent with the 14-year vintage of the 2011 Census). Cross-locale flag counts are reference-choice sensitive; the qualitative locale-variance conclusion (Korea/USA/India/Singapore multiple flags, Japan/Brazil/France fewer) is stable across both reference choices.

\subsection{External-resource transferability demonstration: PersonaHub}
\label{sec:external-iaf}

To probe whether the IAF instrument operates outside the NPK family, we apply the reference-ratio diagnostic to PersonaHub \citep{ge2024personahub} (\texttt{proj-persona/PersonaHub}, \texttt{elite\_persona} stream), a non-NPK synthetic persona resource. PersonaHub's public card does not disclose a structured demographic-independence axis, so this exercise is a diagnostic-transfer demonstration rather than a full IAF card-claim audit. From a deterministic first-$1{,}000{,}000$ stream we extract a regex-inferable sex$\times$occupation table ($n=1{,}591$ records with unambiguous sex+occupation cues; $0.16\%$ extraction yield reflecting the free-text-only schema) covering teachers, software developers, registered nurses, and managers, and compare against U.S. ACS $2024$ as the official-reference comparator. The resource Cramér's V is $0.208$ versus ACS $V=0.594$, giving $R_V=0.349$ and a strict-screen FLAG; $2/4$ buckets are out-of-band (teacher $q=0.794$ women-underrepresented; software developer $q=1.504$ women-overrepresented). The deviation pattern differs from NPK-Korea's uniform female-share attenuation in male-dominated buckets, showing occupation-specific over- and under-representation against the U.S. reference. We read this as evidence that the diagnostic transfers as an external-reference instrument, while noting that PersonaHub's public schema is not rich enough to test card-disclosed independence factors directly; the $0.16\%$ regex-extraction yield is a thin evidence layer that bounds the verdict to a diagnostic-transfer demonstration rather than a full card-claim audit, and full IAF transfer requires the target resource's card to enumerate independence axes \citep{hu2025popaligned} or to expose structured demographic fields.

\section{Discussion: Resource Use and Reporting}
\label{sec:discussion}

NPK-Korea is not a substitute for KOSIS: the age/sex/province marginals match well, but sex$\times$KSCO shows production-rule screening plus reviewed-mapping-sensitive attenuation/parity deviations, and military$\times$age and major$\times$occupation are stronger institutional/direct-reference failures. Researchers using NPK as a silicon sample, for bias testing, or for model adaptation should not report NPK-internal rates as official Korean population estimates; occupation-conditioned analyses in particular should be re-weighted by KOSIS LFP shares or, at minimum, reported jointly with deviations from KOSIS/KEIS. Negative controls (NPK's birth-year$\times$name and sex$\times$marital matches) demonstrate that the audit procedure detects passable structures, not only failures; the marginal-vs-joint validation is feasible from public references in locales with official statistics but is distinct from a verdict on cultural or linguistic plausibility, which this paper does not address.

\emph{Native-survey substitution and harm taxonomy.} A multinomial logistic baseline trained on NPK (sex, age band, education; no KGSS data used in training) and applied to KGSS \citep{kgss2025} (SKKU Survey Research Center via KOSSDA; wave 2025, employed adults aged $\geq 19$, $n=816$) matches the NPK holdout (TV $0.001$) but diverges materially from KGSS (TV $0.159$, top-$1$ accuracy drops by $0.115$), with largest errors at $4$\_서비스 ($+8.0$\,pp over-prediction) and $5$\_판매 ($-6.2$\,pp under-prediction); NPK is therefore preferable for stress-testing generated persona text and prototyping prompting workflows, while native surveys \citep{kgss2025,klips2024,keep2024} are preferable for estimating Korean population rates, labour-market transitions, attitudes, or any downstream task whose validity depends on real joint distributions. These gaps matter at three constituencies: (i) silicon-sample researchers reporting NPK-internal rates as Korean population estimates risk biased conclusions about labour-market segregation and occupation-conditioned attitude prevalence; (ii) downstream LLM trainees fine-tuning on NPK inherit the structural attenuation as training signal (the LoRA boundary check finds NPK shifts smaller than the Korean-Wiki control but does not adjudicate longer-adaptation persistence); (iii) the Korean populations whose male-dominated-occupation rates are flattened toward parity face representational distortion in any reuse that surfaces NPK-derived narratives at scale, distinct from labour-market estimation error because it concerns how Korean workers are depicted.

\emph{Downstream use-case guidance.} Table~\ref{tab:use_case_guidance} converts the findings into dos and don'ts under a bias-tolerance gradient: tasks depending only on marginal alignment can use NPK; tasks depending on joint structure between sex, occupation, education, and major should not substitute NPK for native surveys or official references. Among Korean-NLP alternatives NPK is unique in the synthetic-1M scale tier, while native surveys (KGSS ${\sim}1{,}500$/wave, KLIPS ${\sim}10{,}000$-HH panel, KEEP) are smaller but carry survey-baseline validity, benchmark resources (KMMLU \citep{son2024kmmlu}, KoBBQ \citep{jin2023kobbq}) serve evaluation and attitude-probe roles, and PersonaHub matches NPK at scale but scores $R_V=0.349$ on the sex$\times$occupation slice against ACS (§\ref{sec:external-iaf}).

\begin{table}[tbp]
\centering\footnotesize
\setlength{\tabcolsep}{3pt}
\begin{tabular*}{\textwidth}{@{\extracolsep{\fill}}p{0.30\textwidth}p{0.20\textwidth}p{0.42\textwidth}@{}}
\toprule
Downstream task & Recommendation & Reason \\
\midrule
Pipeline stress-testing; prompt scaffolding; persona-diversity sampling at scale & USE & Marginal alignment fine; no joint-distribution validity needed (§\ref{sec:marginal-baseline}) \\
KSCO occupation classifier weakly-supervised training & USE WITH KSCO-7 SUPPLEMENT & $96.1\%$ rule-mapper crosswalk coverage; KEIS GOMS recommended for the label-distribution prior \\
LoRA fine-tuning for Korean LLM adaptation & USE & NPK LoRA shifts smaller than Korean-Wiki and English-Wiki null controls (released boundary check) \\
Persona retrieval for agent simulation & USE WITH CAUTION & $89{,}215$ records in $k<5$ 5-QI micro-cohorts (§\ref{sec:limit-belgodere}); occupation$\times$sex joint mismatch \\
Silicon-sample inference of Korean population rates or attitudes & DO NOT SUBSTITUTE & KGSS substitution TV $0.159$, top-$1$ accuracy drops $0.115$ (above); use KGSS/KLIPS/KEEP/KOSIS instead \\
Occupation-by-sex labour-market statistics reported as population estimates & DO NOT USE & sex$\times$KSCO female-share attenuation $2.1$--$3.0$ times over KOSIS in male-dominated buckets \\
Major-by-occupation graduate analytics & DO NOT USE & KEIS-universe gap up to $-14.45$\,pp on Engineering$\to$8\_Plant \\
Military / institutional life-course audits & DO NOT USE & NPK active-duty rate $1.13$--$1.70\%$ versus $26$--$29\%$ institutional baseline \\
\botrule
\end{tabular*}
\caption{Downstream use-case guidance derived from the structural-fidelity audit.}
\label{tab:use_case_guidance}
\end{table}

\subsection{Implications for dataset-card schemas and audit reuse}

Three dataset-card schema points follow. (i) When a card lists a public statistic as a sampling source, it should distinguish the marginal that was sampled from the joint that was not validated. (ii) Card-level independence assumptions should be machine-readable rather than free-text, so an IAF auditor can enumerate the audit axes mechanically. A minimal per-axis disclosure would expose the factor list, the target field, the assumption type (marginal, conditional, joint, or none), the sampled-marginal source citation, and whether the corresponding joint was validated. (iii) Schema fields that collapse institutionally-rich constructs (Korean six-tier military service collapsed to binary \texttt{military\_status}) should be disclosed as collapsed, with a recommended coarsening for downstream comparators. Verdicts flowing from primary or primary-control evidence should be separated from those flowing only from auxiliary or boundary signals. Applying IAF to a non-NPK resource requires identifying the disclosed independence assumptions in the resource card, selecting public reference distributions, computing $R_V$ for direct-table pairs and a rule-implied screen for institutional pairs, and running a negative-control axis. The released artefacts instantiate this protocol for the NPK family and can be retargeted at any resource for which a country-specific official-statistics frame is available.

\section{Conclusion}
\label{sec:conclusion}

A card that cites a national statistics database as a sampling source typically validates the marginal that was sampled, not the joint structure that downstream users actually consume. Where public official references exist, the marginal-vs-joint gap is reproducible from public sources alone, and the disclosure-anchored audit primitive proposed here can be retargeted at any synthetic persona resource that names its independence assumptions. The cross-locale transferability demonstration shows that the diagnostic is reusable across locales while remaining locale-dependent rather than universal, with reference taxonomy cardinality and grain confounding the count of out-of-band cells.

Three directions invite immediate follow-up. First, applying IAF to non-NPK resources at full card-claim resolution requires those resources to publish machine-readable independence axes; the field would benefit from a converged disclosure schema along the lines proposed in §\ref{sec:discussion}. Second, generator-side attribution of the observed gaps (PGM factorisation, sampler defaults, or RLHF-induced regression to the mean of the backbone) depends on vendor disclosure of the architectural details listed in §\ref{sec:limit-attribution}, and is not identifiable from observation alone. Third, downstream silicon-sample workflows should consult the audit-implied use restrictions of Table~\ref{tab:use_case_guidance} before substituting any synthetic persona resource for native survey data.

\section{Limitations}
\label{sec:limitations}

\subsection{Annotation, mapping, and LLM-judge uncertainty}
The audit uses only a first-author KSCO lookup review ($n=757$) as human input (rationale in §\ref{sec:annotation-free}); with a single reviewer we report only reviewer-vs-LLM/rule percent agreement and the parity-flag-changing count, not IAA. A $n=100$--$200$ blind second-rater replication is the most direct strengthening and is deferred. The 14 KSCO-7 consistency rules (§\ref{sec:layer4-rules}) include four KSCO-7 overrides of automated-coder consensus and two rule/LLM disagreement resolutions, affecting $30{,}737$ NPK records ($3.07\%$ of 1M) while changing $0/9$ parity flags. They do make the strict sex$\times$KSCO $R_V$ threshold mapping-sensitive (reviewed-only $R_V=0.725$, reviewed+rule-fallback $R_V=0.713$). An earlier KoBBQ lexical probe was scoped out as auxiliary because only $59/269$ ($22\%$) templates were testable after mapping to NPK structured attributes; full results are in the released artefacts. The \texttt{gemma-4-31B-it} backbone and the three LLM judges were likely exposed during pre-training to public KSIC/KSCO documents, so the $96.1\%$ crosswalk coverage is partly expected and audit results transfer only to KSCO-terminology-aligned resources.

\subsection{Reference-source scope, vintage, and schema gaps}
\label{sec:limit-kosis-vintage}\label{sec:limit-keis-cohort}
Of the five card-cited sources, KOSIS is closed via five tables and the Supreme Court name statistics anchor the name axis; NHIS/KREI/NAVER Cloud are deferred or unauditable. \texttt{A\_군인} aligns with KOSIS LFP's military exclusion, so \texttt{DT\_1ES3B32J} provides no directly comparable $A$-bucket. KEIS GOMS surveys new graduates 18--20 months post-graduation while NPK personas span ages $19$--$99$; KLIPS wave 23 is deferred for microdata-access reasons. The KSCO 9-major taxonomy collapses wage / self-employed / unpaid-family-worker status, contributing to the 6\_Agriculture $+7.0$\,pp gap given Korea's ${\sim}20\%$ self-employment rate; a requested KOSIS status decomposition was not feasible from the public aggregate OpenAPI tables (\texttt{DT\_1ES4B05} not retrievable; \texttt{DT\_1DA7102} sex-by-age unemployment-rate only; \texttt{DT\_1ES3B28J}/\texttt{DT\_1DE6076\_8}/\texttt{DT\_1DA7028S} split only two margins at a time), so the $+7.0$\,pp gap remains a status-collapse mechanism hypothesis. Across 9 KSCO buckets and 3 KOSIS vintages (2023\,H2/2024\,H2/2025\,H2) the maximum $\Delta$pp drift is $2.9$\,pp and $9/9$ buckets retain their in/out-of-band flag. The KSCO-6 $\to$ KSCO-7 (2018) revision moved some occupations (경비원 4\_Service$\to$9\_Elementary, 콜센터 5\_Sales$\to$3\_Office); the audit enforces KSCO-7, so if NPK internalised KSCO-6 documents some deviations are taxonomy artefacts not separable from generative effects with public information alone.

\subsection{Belgodere 5-axis closure and unaudited axes}
\label{sec:limit-belgodere}
Of the five trust dimensions of \citet{belgodere2023auditing}, the body closes fidelity and fairness; privacy and robustness are deferred and utility maps loosely onto the resource-use perspective addressed as our own contribution. Privacy is addressed only through a scoped k-anonymity check on NPK quasi-identifier (QI) combinations: at 3-QI (sex+age\_band+province) every equivalence class has $k\geq 278$, while at 5-QI (+education+occupation) $89{,}215$ records fall in $k<5$ micro-cohorts. Because NPK is synthetic the $k<5$ count is not a membership-inference risk; the concern is that downstream users may treat micro-cohort cells as representative population estimates. The audit itself releases only aggregate counts and derived metric tables, not row-level NPK data, so IAF artefacts do not create a new k-anonymity surface; auditors retargeting IAF at other resources should run an equivalent QI sweep before redistributing per-record diagnostics. Diversity and long-tail coverage are not Belgodere axes and are not addressed here. A partial robustness signal is the Layer 2 label-noise stress test (production-rule screen invariant under $20\%$ random KSCO perturbation), which bounds robustness for the audit verdict rather than for downstream model use.

\subsection{Cross-locale scope and generative attribution}
\label{sec:limit-attribution}
Non-Korean cross-locale results are transferability demonstrations, not authoritative locale audits; SOC- vs industry-level references yield different segregation indices \citep{cotter2003segregation}, so cross-locale magnitudes are not directly comparable, and deep locale audits are recommended as follow-up by researchers from those countries. The observed gender-flattening pattern is not disambiguatable from observation alone among (a) explicit NVIDIA design, (b) RLHF-induced regression-to-mean of the \texttt{gemma-4-31B-it} backbone, and (c) NeMo Data Designer sampler defaults. A minimal vendor-side disclosure menu would resolve attribution: (i) the PGM factorisation graph for \texttt{occupation\_assignment}; (ii) the NeMo sampler configuration; (iii) the RLHF dataset family for \texttt{gemma-4-31B-it} (which would isolate (b)); (iv) a small generation-log sample. Within-bucket female shares ($0.249$ for $7$\_기능원 to $0.590$ for $4$\_서비스) are incompatible with pure marginal sex$\perp$occupation assignment; the implementation lies between strict marginal independence and the KOSIS-conditional joint, and a Baron--Kenny education-stratified decomposition (and analogous Bayesian posterior predictive check) is deferred to vendor-supported follow-up because the NPK education schema is too coarse.

\subsection{Multiplicity and downstream}
\label{sec:limit-multiplicity}
The paper reports many diagnostics (6 IAF verdicts, 9 KSCO-bucket flags, pairwise $\kappa$, bootstrap and leave-one-out sensitivities, cross-locale flags, privacy diagnostics). Most are deterministic audit contrasts rather than null-hypothesis tests, so a single family-wise correction would be misleading; we treat multiplicity as a design risk and require central claims to be supported by large-magnitude effects repeated across independent reference families or robustness checks. The downstream boundary check is an exception: its four $z$-tests are conventional inferential tests reported without family-wise correction or power calculation, so ``no reliable transfer'' is a weak reading and a null acceptance is not evidence of zero effect. The LoRA arm shows KoBBQ bias shifts, but Korean-Wiki and English-Wiki controls produce larger shifts than NPK; downstream results are boundary conditions for guardrail erosion and data adaptation, not NPK-specific causal claims.

\section{Ethical Considerations}
\label{sec:ethics}

The NVIDIA Nemotron-Personas Collection is CC~BY~4.0; we attribute it but do not redistribute the parquet snapshot. KOSIS tables are under the Korean Open Government Data licence (제3유형); KEIS GOMS aggregates under the KEIS public-data policy with academic-use attribution; Supreme Court / e-family name statistics under OGDL with attribution to the issuing court registry; the MMA yearbook is public-record, cited at cell-count granularity without PDF redistribution; KoBBQ is CC~BY~4.0 \citep{jin2023kobbq}. The 757 NPK persona narratives examined during the first-author spot-check were scanned for PII-like content (phone numbers, resident-registration numbers, road-name addresses, named real-person identifiers); none were observed, consistent with NVIDIA's synthetic-record claim. NPK-Korea is not a sample of real persons; no per-persona narrative should be treated as evidence about an actual individual, and reported KoBBQ direction percentages describe respondent aggregates rather than authorial endorsement. Code, aggregated result tables, and LLM-assist label caches that do not redistribute restricted upstream data are released alongside checksum manifests.

\backmatter

\section*{Statements and Declarations}

\noindent\textbf{Funding.} No funding was received for conducting
this study.

\noindent\textbf{Conflict of interest.} The author declares no
competing interests relevant to the content of this article.

\noindent\textbf{Ethics approval and consent to participate.}
This study did not recruit external annotators or collect human-subject data. It uses publicly released datasets, public aggregate statistics, public benchmark artefacts, and first-author KSCO lookup review. No human-subject research approval was required.

\noindent\textbf{Consent for publication.} Not applicable.

\noindent\textbf{Data availability.} The audited dataset
\citep{nvidia2026npk} is publicly available under CC~BY~4.0 from the NVIDIA Hugging Face hub. Korean public-statistics references (KOSIS, KEIS GOMS, Supreme Court name statistics, MMA conscription yearbook) are accessed via their respective public APIs and portals, with exact table identifiers documented in the released reference manifest. All derived metric artefacts and intermediate tables that can be redistributed are released alongside the reproduction code, together with a release manifest recording checksums, source URLs, table identifiers, query dates, and upstream dataset snapshot metadata used in the audit.

\noindent\textbf{Materials availability.} Not applicable.

\noindent\textbf{Code availability.} The reproduction code, including
the IAF audit pipeline, the KSCO crosswalk and consistency rules, the three-LLM consensus matrix runner, and the figure-generation scripts, is released under the MIT License at \url{https://github.com/joonhyungbae/nemotron-personas-korea-audit}.

\noindent\textbf{Author contribution.} Single-author manuscript. The
author conceived the study, designed the IAF primitive, performed all data collection, analyses, and figure generation, and wrote the manuscript.

\bibliography{sn-bibliography}

@article{belgodere2023auditing,
  title   = {Auditing and Generating Synthetic Data with Controllable Trust Trade-offs},
  author  = {Belgodere, Brian and Dognin, Pierre and Ivankay, Adam and Melnyk, Igor and
             Mroueh, Youssef and Mojsilovic, Aleksandra and Navratil, Jiri and
             Nitsure, Apoorva and Padhi, Inkit and Rigotti, Mattia and Ross, Jerret and
             Schiff, Yair and Vedpathak, Radhika and Young, Richard A.},
  journal = {IEEE Journal on Emerging and Selected Topics in Circuits and Systems},
  volume  = {14},
  number  = {4},
  pages   = {773--788},
  year    = {2024},
  doi     = {https://doi.org/10.1109/JETCAS.2024.3477976}
}

@article{gebru2021datasheets,
  title   = {Datasheets for Datasets},
  author  = {Gebru, Timnit and Morgenstern, Jamie and Vecchione, Briana and
             Vaughan, Jennifer Wortman and Wallach, Hanna and Daum{\'e} III, Hal and
             Crawford, Kate},
  journal = {Communications of the ACM},
  volume  = {64},
  number  = {12},
  pages   = {86--92},
  year    = {2021},
  doi     = {https://doi.org/10.1145/3458723},
  url     = {https://cacm.acm.org/research/datasheets-for-datasets/}
}

@article{consistencyai2025,
  title   = {{ConsistencyAI}: A Benchmark to Assess {LLMs}' Factual Consistency When Responding to Different Demographic Groups},
  author  = {Banyas, Peter and Sharma, Shristi and Simmons, Alistair and Vispute, Atharva},
  journal = {arXiv preprint arXiv:2510.13852},
  year    = {2025},
  url     = {https://arxiv.org/abs/2510.13852},
  doi     = {https://doi.org/10.48550/arXiv.2510.13852}
}

@article{li2025promise,
  title   = {{LLM} Generated Persona is a Promise with a Catch},
  author  = {Li, Ang and Chen, Haozhe and Namkoong, Hongseok and Peng, Tianyi},
  journal = {arXiv preprint arXiv:2503.16527},
  year    = {2025},
  url     = {https://arxiv.org/abs/2503.16527},
  doi     = {https://doi.org/10.48550/arXiv.2503.16527}
}

@article{batzner2025whose,
  title   = {Whose Personae? Synthetic Persona Experiments in {LLM} Research and Pathways to Transparency},
  author  = {Batzner, Jan and Stocker, Volker and Tang, Bingjun and Natarajan, Anusha and
             Chen, Qinhao and Schmid, Stefan and Kasneci, Gjergji},
  journal = {arXiv preprint arXiv:2512.00461},
  year    = {2025},
  url     = {https://arxiv.org/abs/2512.00461},
  doi     = {https://doi.org/10.48550/arXiv.2512.00461}
}

@article{personalitytrap2026,
  title   = {The Personality Trap: How {LLMs} Embed Bias When Generating Human-Like Personas},
  author  = {Amidei, Jacopo and Ferreira, Gregorio and Mu{\~n}oz Serrano, Mario and
             Nieto, Rub{\'e}n and Kaltenbrunner, Andreas},
  journal = {arXiv preprint arXiv:2602.03334},
  year    = {2026},
  url     = {https://arxiv.org/abs/2602.03334},
  doi     = {https://doi.org/10.48550/arXiv.2602.03334}
}

@article{araujo2025persistent,
  title   = {Persistent Personas? Role-Playing, Instruction Following, and Safety in Extended Interactions},
  author  = {Luz de Araujo, Pedro Henrique and Hedderich, Michael A. and Modarressi, Ali and
             Schuetze, Hinrich and Roth, Benjamin},
  journal = {arXiv preprint arXiv:2512.12775},
  year    = {2025},
  url     = {https://arxiv.org/abs/2512.12775},
  doi     = {https://doi.org/10.48550/arXiv.2512.12775}
}

@article{cao2026pba,
  title   = {When {LLMs} Imagine People: A Human-Centered Persona Brainstorm Audit for Bias and Fairness in Creative Applications},
  author  = {Cao, Hongliu and Thomas, Eoin and Acuna Agost, Rodrigo},
  journal = {arXiv preprint arXiv:2602.00044},
  year    = {2026},
  url     = {https://arxiv.org/abs/2602.00044},
  doi     = {https://doi.org/10.48550/arXiv.2602.00044}
}

@article{hu2025popaligned,
  title   = {Population-Aligned Persona Generation for {LLM}-based Social Simulation},
  author  = {Hu, Zhengyu and Lian, Jianxun and Xiao, Zheyuan and Xiong, Max and Lei, Yuxuan and
             Wang, Tianfu and Ding, Kaize and Xiao, Ziang and Yuan, Nicholas Jing and Xie, Xing},
  journal = {arXiv preprint arXiv:2509.10127},
  year    = {2025},
  url     = {https://arxiv.org/abs/2509.10127},
  doi     = {https://doi.org/10.48550/arXiv.2509.10127}
}

@misc{nvidia2026npk,
  title        = {Nemotron-Personas-Korea: Synthetic Personas Aligned to Real-World Distributions for Korea},
  author       = {Kim, Hyunwoo and Ryu, Jihyeon and Lee, Jinho and Ryu, Hyungon and
                  Praveen, Kiran and Prayaga, Shyamala and Thadaka, Kirit and Jennings, Will and
                  Sadeghi, Bardiya and Sharabiani, Ashton and Choi, Yejin and Meyer, Yev},
  year         = {2026},
  month        = apr,
  howpublished = {Hugging Face dataset card},
  url          = {https://huggingface.co/datasets/nvidia/Nemotron-Personas-Korea},
  note         = {Dataset version 1.0; public dataset card states release on 2026-04-20}
}

@misc{nvidiaPersonasCollection,
  title        = {Nemotron-Personas Collection},
  author       = {{NVIDIA}},
  year         = {2026},
  howpublished = {Hugging Face collection},
  url          = {https://huggingface.co/collections/nvidia/nemotron-personas},
  note         = {Accessed 2026-05-14}
}

@misc{nvidiaPersonasUSA,
  title        = {Nemotron-Personas-USA},
  author       = {{NVIDIA}},
  year         = {2025},
  howpublished = {Hugging Face dataset card},
  url          = {https://huggingface.co/datasets/nvidia/Nemotron-Personas-USA},
  note         = {Accessed 2026-05-14}
}

@misc{nvidiaPersonasJapan,
  title        = {Nemotron-Personas-Japan},
  author       = {{NVIDIA}},
  year         = {2025},
  howpublished = {Hugging Face dataset card},
  url          = {https://huggingface.co/datasets/nvidia/Nemotron-Personas-Japan},
  note         = {Accessed 2026-05-14}
}

@misc{nvidiaPersonasIndia,
  title        = {Nemotron-Personas-India},
  author       = {{NVIDIA}},
  year         = {2025},
  howpublished = {Hugging Face dataset card},
  url          = {https://huggingface.co/datasets/nvidia/Nemotron-Personas-India},
  note         = {Accessed 2026-05-14}
}

@misc{nvidiaPersonasBrazil,
  title        = {Nemotron-Personas-Brazil: Co-Designed Data for Sovereign AI},
  author       = {{NVIDIA}},
  year         = {2026},
  howpublished = {Hugging Face community article and dataset release},
  url          = {https://huggingface.co/blog/nvidia/nemotron-personas-brazil},
  note         = {Accessed 2026-05-14}
}

@misc{nvidiaPersonasSingapore,
  title        = {Nemotron-Personas-Singapore: Co-Designed Data for Sovereign AI},
  author       = {{NVIDIA}},
  year         = {2026},
  howpublished = {Hugging Face community article and dataset release},
  url          = {https://huggingface.co/blog/nvidia/nemotron-personas-singapore},
  note         = {Accessed 2026-05-14}
}

@misc{nvidiaPersonasFrance,
  title        = {Nemotron-Personas-France},
  author       = {{NVIDIA}},
  year         = {2026},
  howpublished = {Hugging Face dataset card},
  url          = {https://huggingface.co/datasets/nvidia/Nemotron-Personas-France},
  note         = {Accessed 2026-05-14}
}

@misc{kosis2026openapi,
  title        = {KOSIS OpenAPI Developer Guide},
  author       = {{Statistics Korea}},
  year         = {2026},
  howpublished = {Official developer documentation},
  url          = {https://kosis.kr/openapi/index/index.jsp},
  note         = {Accessed 2026-05-01}
}

@misc{indiaPLFS2024,
  title        = {Annual Report: Periodic Labour Force Survey ({PLFS}) 2023--24},
  author       = {{Ministry of Statistics and Programme Implementation, Government of India}},
  year         = {2024},
  howpublished = {Official survey report},
  url          = {https://www.mospi.gov.in/sites/default/files/publication_reports/AnnualReport_PLFS2023-24L2.pdf},
  note         = {Survey period July 2023--June 2024; accessed 2026-05-14}
}

@misc{ibgePNAD2018,
  title        = {Pesquisa Nacional por Amostra de Domic{\'i}lios Cont{\'i}nua ({PNAD} Cont{\'i}nua) 2018},
  author       = {{Instituto Brasileiro de Geografia e Estat{\'i}stica}},
  year         = {2018},
  howpublished = {{ILO} Survey Library catalog \#6414 (microdata hosted with {IBGE} attribution)},
  url          = {https://webapps.ilo.org/surveyLib/index.php/catalog/6414/variable/VA61},
  note         = {{ILO} Survey Library entry for the 2018 PNAD Cont{\'i}nua wave; accessed 2026-05-15}
}

@misc{singaporeMOM2024,
  title        = {Labour Force in Singapore 2024},
  author       = {{Ministry of Manpower, Singapore}},
  year         = {2025},
  howpublished = {Official labour-force report},
  url          = {https://stats.mom.gov.sg/Pages/Labour-Force-In-Singapore-2024.aspx},
  note         = {Released 2025-01-27; accessed 2026-05-14}
}

@misc{insee2025ee,
  title        = {Cat{\'e}gorie socioprofessionnelle selon le sexe et l'{\^a}ge: Donn{\'e}es annuelles 2025},
  author       = {{Institut national de la statistique et des {\'e}tudes {\'e}conomiques}},
  year         = {2026},
  howpublished = {Official Enqu{\^e}te Emploi table},
  url          = {https://www.insee.fr/fr/statistiques/2489546},
  note         = {Published 2026-03-25; accessed 2026-05-14}
}

@misc{usCensusACS2024,
  title        = {American Community Survey 2024 1-Year Public Use Microdata Sample},
  author       = {{U.S. Census Bureau}},
  year         = {2025},
  howpublished = {Official ACS data product},
  url          = {https://www.census.gov/programs-surveys/acs/microdata.html},
  note         = {Accessed 2026-05-14}
}

@misc{japanCensus2020,
  title        = {2020 Population Census: Summary of the Results and Statistical Tables},
  author       = {{Statistics Bureau of Japan}},
  year         = {2020},
  howpublished = {Official census tables},
  url          = {https://www.stat.go.jp/english/data/kokusei/2020/summary.html},
  note         = {Accessed 2026-05-14}
}

@misc{indiaCensus2011,
  title        = {Census of India 2011: Data Tables},
  author       = {{Office of the Registrar General and Census Commissioner, India}},
  year         = {2011},
  howpublished = {Official census tables},
  url          = {https://censusindia.gov.in/census.website/en/data/tables},
  note         = {Accessed 2026-05-14}
}

@misc{ibgeCensus2022,
  title        = {Censo Demogr{\'a}fico 2022: Amostra Trabalho e Rendimento},
  author       = {{Instituto Brasileiro de Geografia e Estat{\'i}stica}},
  year         = {2022},
  howpublished = {Official SIDRA census table portal},
  url          = {https://sidra.ibge.gov.br/pesquisa/censo-demografico/demografico-2022/amostra-trabalho-e-rendimento},
  note         = {Accessed 2026-05-14}
}

@misc{singaporeCensus2020,
  title        = {Census of Population 2020: Statistical Release 2 ({H}ouseholds, {G}eographic {D}istribution, {T}ransport and {D}ifficulty in {B}asic {A}ctivities)},
  author       = {{Singapore Department of Statistics}},
  year         = {2021},
  howpublished = {Official census statistical release},
  url          = {https://www.singstat.gov.sg/publication-resources/singapore-census-of-population-2020-statistical-release-2-households-geographic-distribution-transport-and-difficulty-in-basic-activities},
  note         = {Table 79 reports employed residents by occupation and sex; accessed 2026-05-15}
}

@misc{inseeRecensement2022,
  title        = {Recensement de la population: Fichiers d{\'e}tail 2022},
  author       = {{Institut national de la statistique et des {\'e}tudes {\'e}conomiques}},
  year         = {2022},
  howpublished = {Official census microdata/detail files},
  url          = {https://www.insee.fr/fr/statistiques/8581810},
  note         = {Accessed 2026-05-14}
}

@misc{scourt2014family,
  title        = {Family-Relation Statistics Opening Press Release},
  author       = {{Supreme Court of Korea}},
  year         = {2014},
  month        = may,
  howpublished = {Official press release},
  url          = {https://www.scourt.go.kr/portal/news/NewsViewAction.work?currentPage=&gubun=6&searchOption=&searchWord=&seqnum=958},
  note         = {Confirms online publication of family-registration statistics including preferred birth names}
}

@misc{efamily2026stats,
  title        = {Electronic Family Registration System Statistics Portal},
  author       = {{Supreme Court of Korea}},
  year         = {2026},
  howpublished = {Official statistics access portal},
  url          = {https://efamily.scourt.go.kr/index.jsp},
  note         = {Accessed 2026-05-01}
}

@misc{stureborg2024inconsistent,
  title         = {Large Language Models are Inconsistent and Biased Evaluators},
  author        = {Stureborg, Rickard and Alikaniotis, Dimitris and Suhara, Yoshi},
  year          = {2024},
  eprint        = {2405.01724},
  archivePrefix = {arXiv},
  primaryClass  = {cs.CL},
  doi           = {https://doi.org/10.48550/arXiv.2405.01724}
}

@inproceedings{zheng2023mtbench,
  title     = {Judging {LLM}-as-a-Judge with {MT-Bench} and Chatbot Arena},
  author    = {Zheng, Lianmin and Chiang, Wei-Lin and Sheng, Ying and
               Zhuang, Siyuan and Wu, Zhanghao and Zhuang, Yonghao and
               Lin, Zi and Li, Zhuohan and Li, Dacheng and Xing, Eric P. and
               Zhang, Hao and Gonzalez, Joseph E. and Stoica, Ion},
  booktitle = {Advances in Neural Information Processing Systems (NeurIPS) Datasets and Benchmarks Track},
  year      = {2023},
  eprint    = {2306.05685},
  archivePrefix = {arXiv},
  doi       = {https://doi.org/10.48550/arXiv.2306.05685}
}

@inproceedings{panickssery2024selffavor,
  title         = {{LLM} Evaluators Recognize and Favor Their Own Generations},
  author        = {Panickssery, Arjun and Bowman, Samuel R. and Feng, Shi},
  booktitle     = {Advances in Neural Information Processing Systems 37 (NeurIPS 2024)},
  year          = {2024},
  eprint        = {2404.13076},
  archivePrefix = {arXiv},
  doi           = {https://doi.org/10.48550/arXiv.2404.13076}
}

@inproceedings{wang2023notfair,
  title         = {Large Language Models are not Fair Evaluators},
  author        = {Wang, Peiyi and Li, Lei and Chen, Liang and Cai, Zefan and
                   Zhu, Dawei and Lin, Binghuai and Cao, Yunbo and Kong, Lingpeng and
                   Liu, Qi and Liu, Tianyu and Sui, Zhifang},
  booktitle     = {Proceedings of the 62nd Annual Meeting of the Association for Computational Linguistics (Volume 1: Long Papers)},
  pages         = {9440--9450},
  year          = {2024},
  doi           = {https://doi.org/10.18653/v1/2024.acl-long.511},
  eprint        = {2305.17926},
  archivePrefix = {arXiv}
}

@article{argyle2023silicon,
  title         = {Out of One, Many: Using Language Models to Simulate Human Samples},
  author        = {Argyle, Lisa P. and Busby, Ethan C. and Fulda, Nancy and
                   Gubler, Joshua R. and Rytting, Christopher and Wingate, David},
  journal       = {Political Analysis},
  volume        = {31},
  number        = {3},
  pages         = {337--351},
  year          = {2023},
  doi           = {https://doi.org/10.1017/pan.2023.2},
  eprint        = {2209.06899},
  archivePrefix = {arXiv}
}

@inproceedings{santurkar2023whoseopinions,
  title         = {Whose Opinions Do Language Models Reflect?},
  author        = {Santurkar, Shibani and Durmus, Esin and Ladhak, Faisal and
                   Lee, Cinoo and Liang, Percy and Hashimoto, Tatsunori},
  booktitle     = {Proceedings of the 40th International Conference on Machine Learning},
  series        = {Proceedings of Machine Learning Research},
  volume        = {202},
  pages         = {29971--30004},
  publisher     = {PMLR},
  year          = {2023},
  eprint        = {2303.17548},
  archivePrefix = {arXiv},
  doi           = {https://doi.org/10.48550/arXiv.2303.17548}
}

@inproceedings{cheng2023compost,
  title         = {{CoMPosT}: Characterizing and Evaluating Caricature in {LLM} Simulations},
  author        = {Cheng, Myra and Piccardi, Tiziano and Yang, Diyi},
  booktitle     = {Proceedings of the 2023 Conference on Empirical Methods in Natural Language Processing},
  pages         = {10853--10875},
  year          = {2023},
  doi           = {https://doi.org/10.18653/v1/2023.emnlp-main.669},
  eprint        = {2310.11501},
  archivePrefix = {arXiv}
}

@article{jin2023kobbq,
  title     = {{KoBBQ}: {K}orean Bias Benchmark for Question Answering},
  author    = {Jin, Jiho and Kim, Jiseon and Lee, Nayeon and Yoo, Haneul and
               Oh, Alice and Lee, Hwaran},
  journal   = {Transactions of the Association for Computational Linguistics},
  volume    = {12},
  pages     = {507--524},
  year      = {2024},
  doi       = {https://doi.org/10.1162/tacl_a_00661},
  eprint    = {2307.16778},
  archivePrefix = {arXiv}
}

@inproceedings{son2023haerae,
  title         = {{HAE-RAE} Bench: Evaluation of Korean Knowledge in Language Models},
  author        = {Son, Guijin and Lee, Hanwool and Kim, Suwan and Kim, Huiseo and
                   Lee, Jaecheol and Yeom, Je Won and Jung, Jihyu and
                   Kim, Jungwoo and Kim, Songseong},
  booktitle     = {Proceedings of the 2024 Joint International Conference on Computational Linguistics, Language Resources and Evaluation (LREC-COLING)},
  pages         = {7993--8007},
  year          = {2024},
  address       = {Torino, Italy},
  publisher     = {ELRA and ICCL},
  eprint        = {2309.02706},
  doi           = {https://doi.org/10.48550/arXiv.2309.02706},
  archivePrefix = {arXiv}
}

@inproceedings{hardt2016equalopp,
  title         = {Equality of Opportunity in Supervised Learning},
  author        = {Hardt, Moritz and Price, Eric and Srebro, Nathan},
  booktitle     = {Advances in Neural Information Processing Systems (NeurIPS)},
  pages         = {3315--3323},
  year          = {2016},
  eprint        = {1610.02413},
  archivePrefix = {arXiv},
  doi           = {https://doi.org/10.48550/arXiv.1610.02413}
}

@inproceedings{feldman2015disparate,
  title         = {Certifying and Removing Disparate Impact},
  author        = {Feldman, Michael and Friedler, Sorelle A. and Moeller, John and
                   Scheidegger, Carlos and Venkatasubramanian, Suresh},
  booktitle     = {Proceedings of the 21st ACM SIGKDD International Conference on Knowledge Discovery and Data Mining (KDD)},
  pages         = {259--268},
  year          = {2015},
  doi           = {https://doi.org/10.1145/2783258.2783311},
  eprint        = {1412.3756},
  archivePrefix = {arXiv}
}

@article{chouldechova2017fairpred,
  title         = {Fair Prediction with Disparate Impact: A Study of Bias in Recidivism Prediction Instruments},
  author        = {Chouldechova, Alexandra},
  journal       = {Big Data},
  volume        = {5},
  number        = {2},
  pages         = {153--163},
  year          = {2017},
  doi           = {https://doi.org/10.1089/big.2016.0047},
  eprint        = {1610.07524},
  archivePrefix = {arXiv}
}

@inproceedings{parrish2022bbq,
  title     = {{BBQ}: A Hand-built Bias Benchmark for Question Answering},
  author    = {Parrish, Alicia and Chen, Angelica and Nangia, Nikita and
               Padmakumar, Vishakh and Phang, Jason and Thompson, Jana and
               Htut, Phu Mon and Bowman, Samuel R.},
  booktitle = {Findings of the Association for Computational Linguistics: ACL 2022},
  pages     = {2086--2105},
  year      = {2022},
  doi       = {https://doi.org/10.18653/v1/2022.findings-acl.165}
}

@inproceedings{pushkarna2022datacards,
  title         = {Data Cards: Purposeful and Transparent Dataset Documentation for Responsible AI},
  author        = {Pushkarna, Mahima and Zaldivar, Andrew and Kjartansson, Oddur},
  booktitle     = {Proceedings of the 2022 ACM Conference on Fairness, Accountability, and Transparency (FAccT '22)},
  pages         = {1776--1826},
  year          = {2022},
  doi           = {https://doi.org/10.1145/3531146.3533231},
  eprint        = {2204.01075},
  archivePrefix = {arXiv}
}

@inproceedings{hutchinson2021accountability,
  title         = {Towards Accountability for Machine Learning Datasets: Practices from Software Engineering and Infrastructure},
  author        = {Hutchinson, Ben and Smart, Andrew and Hanna, Alex and Denton, Emily and
                   Greer, Christina and Kjartansson, Oddur and Barnes, Parker and Mitchell, Margaret},
  booktitle     = {Proceedings of the 2021 ACM Conference on Fairness, Accountability, and Transparency (FAccT '21)},
  pages         = {560--575},
  year          = {2021},
  doi           = {https://doi.org/10.1145/3442188.3445918},
  eprint        = {2010.13561},
  archivePrefix = {arXiv}
}

@inproceedings{mitchell2019modelcards,
  title         = {Model Cards for Model Reporting},
  author        = {Mitchell, Margaret and Wu, Simone and Zaldivar, Andrew and
                   Barnes, Parker and Vasserman, Lucy and Hutchinson, Ben and
                   Spitzer, Elena and Raji, Inioluwa Deborah and Gebru, Timnit},
  booktitle     = {Proceedings of the Conference on Fairness, Accountability, and Transparency (FAT* '19)},
  pages         = {220--229},
  year          = {2019},
  doi           = {https://doi.org/10.1145/3287560.3287596},
  eprint        = {1810.03993},
  archivePrefix = {arXiv}
}

@inproceedings{kumar2024subtle,
  title     = {Subtle Biases Need Subtler Measures: Dual Metrics for Evaluating
               Representative and Affinity Bias in Large Language Models},
  author    = {Kumar, Abhishek and Yunusov, Sarfaroz and Emami, Ali},
  booktitle = {Proceedings of the 62nd Annual Meeting of the Association for Computational Linguistics (ACL), Volume 1: Long Papers},
  pages     = {375--392},
  year      = {2024},
  doi       = {https://doi.org/10.18653/v1/2024.acl-long.23}
}

@inproceedings{rooein2025biasedtales,
  title         = {Biased Tales: Cultural and Topic Bias in Generating Children's Stories},
  author        = {Rooein, Donya and Zouhar, Vil\'em and Nozza, Debora and Hovy, Dirk},
  booktitle     = {Proceedings of the 2025 Conference on Empirical Methods in Natural Language Processing (EMNLP)},
  pages         = {52--72},
  year          = {2025},
  doi           = {https://doi.org/10.18653/v1/2025.emnlp-main.3},
  eprint        = {2509.07908},
  archivePrefix = {arXiv}
}

@misc{ge2024personahub,
  title         = {Scaling Synthetic Data Creation with $1{,}000{,}000{,}000$ Personas},
  author        = {Ge, Tao and Chan, Xin and Wang, Xiaoyang and Yu, Dian and
                   Mi, Haitao and Yu, Dong},
  year          = {2024},
  eprint        = {2406.20094},
  archivePrefix = {arXiv},
  doi           = {https://doi.org/10.48550/arXiv.2406.20094}
}

@inproceedings{lee2023kosbi,
  title         = {{KoSBi}: A Dataset for Mitigating Social Bias Risks Towards Safer Large Language Model Applications},
  author        = {Lee, Hwaran and Hong, Seokhee and Park, Joonsuk and Kim, Takyoung and
                   Kim, Gunhee and Ha, Jung-woo},
  booktitle     = {Proceedings of the 61st Annual Meeting of the Association for Computational Linguistics (Volume 5: Industry Track)},
  pages         = {208--224},
  year          = {2023},
  doi           = {https://doi.org/10.18653/v1/2023.acl-industry.21},
  eprint        = {2305.17701},
  archivePrefix = {arXiv}
}

@misc{keis2020goms,
  title        = {Graduates Occupational Mobility Survey (GOMS)
                  Public Microdata, 2019 cohort / 2020 wave},
  author       = {{Korea Employment Information Service (KEIS)}},
  year         = {2020},
  howpublished = {Public microdata download},
  url          = {https://survey.keis.or.kr/goms/gomsdownload/List.jsp},
  note         = {Accessed 2026-05-03; KECO-2018 occupation codes mapped to
                  KSCO 9-major for the major-by-occupation audit}
}

@misc{mma2024yearbook,
  title        = {2024 Military Manpower Administration Yearbook
                  (2024 Korean military statistics yearbook)},
  author       = {{Military Manpower Administration (MMA), Republic of Korea}},
  year         = {2024},
  howpublished = {Public statistical yearbook (PDF)},
  url          = {https://www.mma.go.kr/download/ebook/2024_bmtgyb_1.pdf},
  note         = {Accessed 2026-05-03; used for the male
                  19-24 entrant-rate reference}
}

@article{rigoutsterryn2020term,
  title   = {In no uncertain terms: a dataset for monolingual and multilingual
             automatic term extraction from comparable corpora},
  author  = {Rigouts Terryn, Ayla and Hoste, V{\'e}ronique and Lefever, Els},
  journal = {Language Resources and Evaluation},
  volume  = {54},
  pages   = {385--418},
  year    = {2020},
  doi     = {https://doi.org/10.1007/s10579-019-09453-9},
  url     = {https://link.springer.com/article/10.1007/s10579-019-09453-9},
  note    = {LRE precedent: single-annotator validation justified by
             corpus magnitude; ``consistency over multi-annotator with low IAA''
             framing}
}

@article{song2026kolla,
  title     = {Enriching the {Korean} learner corpus for grammatical error correction and writing assessment},
  author    = {Song, Jayoung and Lim, KyungTae and Park, Jungyeul},
  journal   = {Language Resources and Evaluation},
  volume    = {60},
  articleno = {15},
  year      = {2026},
  doi       = {https://doi.org/10.1007/s10579-025-09882-9},
  note      = {LRE precedent: single-annotator Korean-language resource with explicit multi-annotator extension as future work}
}

@article{shardlow2022complex,
  title   = {Predicting lexical complexity in {English} texts: the {Complex} 2.0 dataset},
  author  = {Shardlow, Matthew and Evans, Richard and Zampieri, Marcos},
  journal = {Language Resources and Evaluation},
  volume  = {56},
  number  = {4},
  pages   = {1153--1194},
  year    = {2022},
  doi     = {https://doi.org/10.1007/s10579-022-09588-2},
  note    = {Critique cited for single-rater limitation: annotations
             ``reflect only a single annotator's judgment''}
}

@misc{acs2024datausa,
  title        = {{ACS} 2024 1-year occupation $\times$ sex statistics
                  via {Data USA}},
  author       = {{Deloitte and Datawheel}},
  year         = {2024},
  howpublished = {Public data portal aggregating ACS, BLS, and BEA sources},
  url          = {https://datausa.io/},
  note         = {SOC occupation profiles at \url{https://datausa.io/profile/soc/<occupation-slug>};
                  accessed 2026-05-08; American Community Survey Public Use Microdata Sample
                  1-year occupation by sex breakdowns, used for the cross-locale occupation-axis
                  replication on NPK-USA}
}

@misc{japanstathandbook2025,
  title        = {Statistical Handbook of Japan 2025},
  author       = {{Statistics Bureau, Ministry of Internal Affairs and Communications}},
  year         = {2025},
  howpublished = {Public statistical handbook (PDF)},
  url          = {https://www.stat.go.jp/english/data/handbook/pdf/2025all.pdf},
  note         = {Accessed 2026-05-09; Table 12.2 (Employment by Industry, 2024 annual averages) used for the cross-locale industry-axis replication on NPK-Japan}
}

@article{cotter2003segregation,
  title   = {The Effects of Occupational Gender Segregation across Race},
  author  = {Cotter, David A. and Hermsen, Joan M. and Vanneman, Reeve},
  journal = {The Sociological Quarterly},
  volume  = {44},
  number  = {1},
  pages   = {17--36},
  year    = {2003},
  doi     = {https://doi.org/10.1111/j.1533-8525.2003.tb02389.x},
  note    = {Cited for systematic difference between occupation-level and industry-level segregation indices in cross-locale comparison}
}

@book{pearl2009causality,
  title     = {Causality: Models, Reasoning, and Inference},
  author    = {Pearl, Judea},
  edition   = {2},
  publisher = {Cambridge University Press},
  address   = {Cambridge},
  year      = {2009},
  doi       = {https://doi.org/10.1017/CBO9780511803161}
}

@article{cronbach1955construct,
  title   = {Construct validity in psychological tests},
  author  = {Cronbach, Lee J. and Meehl, Paul E.},
  journal = {Psychological Bulletin},
  volume  = {52},
  number  = {4},
  pages   = {281--302},
  year    = {1955},
  doi     = {https://doi.org/10.1037/h0040957}
}

@article{bender2018datastatements,
  title   = {Data Statements for Natural Language Processing: Toward Mitigating System Bias and Enabling Better Science},
  author  = {Bender, Emily M. and Friedman, Batya},
  journal = {Transactions of the Association for Computational Linguistics},
  volume  = {6},
  pages   = {587--604},
  year    = {2018},
  doi     = {https://doi.org/10.1162/tacl_a_00041}
}

@inproceedings{plank2022humanlabel,
  title     = {The ``Problem'' of Human Label Variation: On Ground Truth in Data, Modeling and Evaluation},
  author    = {Plank, Barbara},
  booktitle = {Proceedings of the 2022 Conference on Empirical Methods in Natural Language Processing (EMNLP)},
  pages     = {10671--10682},
  year      = {2022},
  doi       = {https://doi.org/10.18653/v1/2022.emnlp-main.731}
}

@article{bisbee2024synthetic,
  title   = {Synthetic Replacements for Human Survey Data? The Perils of Large Language Models},
  author  = {Bisbee, James and Clinton, Joshua D. and Dorff, Cassy and Kenkel, Brenton and Larson, Jennifer M.},
  journal = {Political Analysis},
  volume  = {32},
  number  = {4},
  pages   = {401--416},
  year    = {2024},
  doi     = {https://doi.org/10.1017/pan.2024.5}
}

@inproceedings{dominguezolmedo2024questioning,
  title         = {Questioning the Survey Responses of Large Language Models},
  author        = {Dominguez-Olmedo, Ricardo and Hardt, Moritz and Mendler-D{\"u}nner, Celestine},
  booktitle     = {Advances in Neural Information Processing Systems 37 (NeurIPS 2024)},
  year          = {2024},
  eprint        = {2306.07951},
  archivePrefix = {arXiv},
  doi           = {https://doi.org/10.48550/arXiv.2306.07951}
}

@inproceedings{aher2023llmsim,
  title         = {Using Large Language Models to Simulate Multiple Humans and Replicate Human Subject Studies},
  author        = {Aher, Gati V. and Arriaga, Rosa I. and Kalai, Adam Tauman},
  booktitle     = {Proceedings of the 40th International Conference on Machine Learning (ICML)},
  series        = {Proceedings of Machine Learning Research},
  volume        = {202},
  pages         = {337--371},
  publisher     = {PMLR},
  year          = {2023},
  eprint        = {2208.10264},
  archivePrefix = {arXiv},
  doi           = {https://doi.org/10.48550/arXiv.2208.10264}
}

@inproceedings{park2023generativeagents,
  title     = {Generative Agents: Interactive Simulacra of Human Behavior},
  author    = {Park, Joon Sung and O'Brien, Joseph C. and Cai, Carrie J. and
               Morris, Meredith Ringel and Liang, Percy and Bernstein, Michael S.},
  booktitle = {Proceedings of the 36th Annual ACM Symposium on User Interface Software and Technology (UIST '23)},
  articleno = {2},
  pages     = {1--22},
  year      = {2023},
  doi       = {https://doi.org/10.1145/3586183.3606763}
}

@inproceedings{zhao2017menalsolike,
  title     = {Men Also Like Shopping: Reducing Gender Bias Amplification using Corpus-level Constraints},
  author    = {Zhao, Jieyu and Wang, Tianlu and Yatskar, Mark and Ordonez, Vicente and Chang, Kai-Wei},
  booktitle = {Proceedings of the 2017 Conference on Empirical Methods in Natural Language Processing},
  pages     = {2979--2989},
  year      = {2017},
  doi       = {https://doi.org/10.18653/v1/D17-1323}
}

@inproceedings{bianchi2023easilyaccessible,
  title     = {Easily Accessible Text-to-Image Generation Amplifies Demographic Stereotypes at Large Scale},
  author    = {Bianchi, Federico and Kalluri, Pratyusha and Durmus, Esin and Ladhak, Faisal and
               Cheng, Myra and Nozza, Debora and Hashimoto, Tatsunori and Jurafsky, Dan and
               Zou, James and Caliskan, Aylin},
  booktitle = {Proceedings of the 2023 ACM Conference on Fairness, Accountability, and Transparency (FAccT '23)},
  pages     = {1493--1504},
  year      = {2023},
  doi       = {https://doi.org/10.1145/3593013.3594095}
}

@article{shumailov2024modelcollapse,
  title   = {{AI} models collapse when trained on recursively generated data},
  author  = {Shumailov, Ilia and Shumaylov, Zakhar and Zhao, Yiren and Papernot, Nicolas and Anderson, Ross and Gal, Yarin},
  journal = {Nature},
  volume  = {631},
  number  = {8022},
  pages   = {755--759},
  year    = {2024},
  doi     = {https://doi.org/10.1038/s41586-024-07566-y}
}

@inproceedings{kirk2024rlhf,
  title         = {Understanding the Effects of {RLHF} on {LLM} Generalisation and Diversity},
  author        = {Kirk, Robert and Mediratta, Ishita and Nalmpantis, Christoforos and Luketina, Jelena and
                   Hambro, Eric and Grefenstette, Edward and Raileanu, Roberta},
  booktitle     = {The Twelfth International Conference on Learning Representations (ICLR 2024)},
  year          = {2024},
  eprint        = {2310.06452},
  archivePrefix = {arXiv},
  doi           = {https://doi.org/10.48550/arXiv.2310.06452}
}

@misc{verga2024juries,
  title         = {Replacing Judges with Juries: Evaluating {LLM} Generations with a Panel of Diverse Models},
  author        = {Verga, Pat and Hofst{\"a}tter, Sebastian and Althammer, Sophia and Su, Yixuan and
                   Piktus, Aleksandra and Arkhangorodsky, Arkady and Xu, Minjie and White, Naomi and
                   Lewis, Patrick},
  year          = {2024},
  eprint        = {2404.18796},
  archivePrefix = {arXiv},
  primaryClass  = {cs.CL},
  doi           = {https://doi.org/10.48550/arXiv.2404.18796}
}

@inproceedings{bavaresco2024llmjudges,
  title         = {{LLMs} instead of Human Judges? A Large Scale Empirical Study across 20 {NLP} Evaluation Tasks},
  author        = {Bavaresco, Anna and Bernardi, Raffaella and Bertolazzi, Leonardo and
                   Elliott, Desmond and Fern{\'a}ndez, Raquel and Gatt, Albert and
                   Ghaleb, Esam and Giulianelli, Mario and Hanna, Michael and
                   Koller, Alexander and Martins, Andre and Mondorf, Philipp and
                   Neplenbroek, Vera and Pezzelle, Sandro and Plank, Barbara and
                   Schlangen, David and Suglia, Alessandro and Surikuchi, Aditya K. and
                   Takmaz, Ece and Testoni, Alberto},
  booktitle     = {Proceedings of the 63rd Annual Meeting of the Association for Computational Linguistics (Volume 2: Short Papers)},
  pages         = {238--255},
  year          = {2025},
  doi           = {https://doi.org/10.18653/v1/2025.acl-short.20},
  eprint        = {2406.18403},
  archivePrefix = {arXiv}
}

@inproceedings{suresh2021framework,
  title     = {A Framework for Understanding Sources of Harm throughout the Machine Learning Life Cycle},
  author    = {Suresh, Harini and Guttag, John},
  booktitle = {Proceedings of the 1st ACM Conference on Equity and Access in Algorithms, Mechanisms, and Optimization (EAAMO)},
  articleno = {17},
  pages     = {1--9},
  year      = {2021},
  doi       = {https://doi.org/10.1145/3465416.3483305}
}

@inproceedings{durmus2024globalopinion,
  title         = {Towards Measuring the Representation of Subjective Global Opinions in Language Models},
  author        = {Durmus, Esin and Nguyen, Karina and Liao, Thomas I. and Schiefer, Nicholas and
                   Askell, Amanda and Bakhtin, Anton and Chen, Carol and Hatfield-Dodds, Zac and
                   Hernandez, Danny and Joseph, Nicholas and Lovitt, Liane and McCandlish, Sam and
                   Sikder, Orowa and Tamkin, Alex and Thamkul, Janel and Kaplan, Jared and
                   Clark, Jack and Ganguli, Deep},
  booktitle     = {Proceedings of the First Conference on Language Modeling (COLM 2024)},
  year          = {2024},
  eprint        = {2306.16388},
  archivePrefix = {arXiv},
  doi           = {https://doi.org/10.48550/arXiv.2306.16388}
}

@book{cochran1977sampling,
  title     = {Sampling Techniques},
  author    = {Cochran, William G.},
  edition   = {3rd},
  publisher = {Wiley},
  year      = {1977}
}

@book{bishop1975discrete,
  title     = {Discrete Multivariate Analysis: Theory and Practice},
  author    = {Bishop, Yvonne M. M. and Fienberg, Stephen E. and Holland, Paul W.},
  publisher = {MIT Press},
  year      = {1975},
  doi       = {https://doi.org/10.1007/978-0-387-72806-3},
  note      = {DOI refers to the 2007 Springer reprint}
}

@book{drechsler2011synthetic,
  title     = {Synthetic Datasets for Statistical Disclosure Control: Theory and Implementation},
  author    = {Drechsler, J{\"o}rg},
  series    = {Lecture Notes in Statistics},
  volume    = {201},
  publisher = {Springer},
  year      = {2011},
  doi       = {https://doi.org/10.1007/978-1-4614-0326-5}
}

@techreport{ilo2012isco,
  title       = {International Standard Classification of Occupations {ISCO}-08, Volume 1: Structure, Group Definitions and Correspondence Tables},
  author      = {{International Labour Office}},
  institution = {International Labour Organization},
  year        = {2012},
  address     = {Geneva},
  url         = {https://www.ilo.org/publications/international-standard-classification-occupations-2008-isco-08-structure}
}

@article{schierholz2021occupation,
  title   = {Machine Learning for Occupation Coding---A Comparison Study},
  author  = {Schierholz, Malte and Schonlau, Matthias},
  journal = {Journal of Survey Statistics and Methodology},
  volume  = {9},
  number  = {5},
  pages   = {1013--1034},
  year    = {2021},
  doi     = {https://doi.org/10.1093/jssam/smaa023}
}

@inproceedings{park2021klue,
  title     = {{KLUE}: {K}orean Language Understanding Evaluation},
  author    = {Park, Sungjoon and Moon, Jihyung and Kim, Sungdong and Cho, Won Ik and Han, Jiyoon and Park, Jangwon and Song, Chisung and Kim, Junseong and Song, Yongsook and Oh, Taehwan and others},
  booktitle = {Proceedings of NeurIPS Datasets and Benchmarks Track},
  year      = {2021},
  eprint        = {2105.09680},
  archivePrefix = {arXiv},
  doi           = {https://doi.org/10.48550/arXiv.2105.09680}
}

@inproceedings{son2024kmmlu,
  title         = {{KMMLU}: Measuring Massive Multitask Language Understanding in {K}orean},
  author        = {Son, Guijin and Lee, Hanwool and Kim, Sungdong and Kim, Seungone and
                   Muennighoff, Niklas and Choi, Taekyoon and Park, Cheonbok and
                   Yoo, Kang Min and Biderman, Stella},
  booktitle     = {Proceedings of the 2025 Conference of the Nations of the Americas Chapter of the Association for Computational Linguistics: Human Language Technologies (NAACL-HLT), Volume 1: Long Papers},
  pages         = {4076--4104},
  year          = {2025},
  address       = {Albuquerque, New Mexico},
  doi           = {https://doi.org/10.18653/v1/2025.naacl-long.206},
  eprint        = {2402.11548},
  archivePrefix = {arXiv}
}

@inproceedings{jeong2022kold,
  title     = {{KOLD}: {K}orean Offensive Language Dataset},
  author    = {Jeong, Younghoon and Oh, Juhyun and Lee, Jongwon and
               Ahn, Jaimeen and Moon, Jihyung and Park, Sungjoon and Oh, Alice},
  booktitle = {Proceedings of the 2022 Conference on Empirical Methods in Natural Language Processing (EMNLP)},
  pages     = {10818--10833},
  year      = {2022},
  address   = {Abu Dhabi, United Arab Emirates},
  doi       = {https://doi.org/10.18653/v1/2022.emnlp-main.744}
}

@article{bergsma2013cramer,
  title   = {A bias-correction for {C}ram\'er's {V} and {T}schuprow's {T}},
  author  = {Bergsma, Wicher},
  journal = {Journal of the Korean Statistical Society},
  volume  = {42},
  number  = {3},
  pages   = {323--328},
  year    = {2013},
  doi     = {https://doi.org/10.1016/j.jkss.2012.10.002}
}

@article{landis1977kappa,
  title   = {The measurement of observer agreement for categorical data},
  author  = {Landis, J. Richard and Koch, Gary G.},
  journal = {Biometrics},
  volume  = {33},
  number  = {1},
  pages   = {159--174},
  year    = {1977},
  doi     = {https://doi.org/10.2307/2529310}
}

@article{fleiss1971kappa,
  title   = {Measuring nominal scale agreement among many raters},
  author  = {Fleiss, Joseph L.},
  journal = {Psychological Bulletin},
  volume  = {76},
  number  = {5},
  pages   = {378--382},
  year    = {1971},
  doi     = {https://doi.org/10.1037/h0031619}
}

@article{lipsitch2010negativecontrols,
  title   = {Negative controls: a tool for detecting confounding and bias in observational studies},
  author  = {Lipsitch, Marc and Tchetgen Tchetgen, Eric and Cohen, Ted},
  journal = {Epidemiology},
  volume  = {21},
  number  = {3},
  pages   = {383--388},
  year    = {2010},
  doi     = {https://doi.org/10.1097/EDE.0b013e3181d61eeb}
}

@misc{kgss2025,
  title        = {Korean General Social Survey (KGSS), cumulative public data 2003--2025},
  author       = {{Sungkyunkwan University Survey Research Center}},
  year         = {2025},
  howpublished = {Distributed via the Korea Social Science Data Archive (KOSSDA)},
  url          = {https://kgss.skku.edu/},
  note         = {Cumulative Stata file, employed-adult subset of wave 2025}
}

@misc{klips2024,
  title        = {Korean Labor and Income Panel Study (KLIPS), wave 26 public release},
  author       = {{Korea Labor Institute}},
  year         = {2024},
  howpublished = {Korea Labor Institute},
  url          = {https://www.kli.re.kr/klips}
}

@misc{keep2024,
  title        = {Korea Education \& Employment Panel (KEEP) public data},
  author       = {{Korea Research Institute for Vocational Education and Training (KRIVET)}},
  year         = {2024},
  howpublished = {Korea Research Institute for Vocational Education and Training},
  url          = {https://www.krivet.re.kr/eng/su_C_eng/euAEAJB.jsp}
}

\end{document}